\documentclass{myacmtrans2m}

\acmVolume{}
\acmNumber{}
\acmYear{}
\acmMonth{}

\usepackage{multirow}
\usepackage{epsfig}

\newcommand{\BibTeX}{{\rm B\kern-.05em{\sc i\kern-.025em b}\kern-.08em
    T\kern-.1667em\lower.7ex\hbox{E}\kern-.125emX}}
    
\title{Survey of Technologies for Web Application Development}

\author{BARRY DOYLE\\University of California, Irvine \and \\
CRISTINA VIDEIRA LOPES\\University of California, Irvine}

\begin{abstract}
Web-based application developers face a dizzying array of platforms,
languages, frameworks and technical artifacts to choose from. We
survey, classify, and compare technologies supporting Web application
development. The classification is based on (1) foundational
technologies; (2) integration with other information sources; and (3)
dynamic content generation. We further survey and classify software
engineering techniques and tools that have been adopted from
traditional programming into Web programming. We conclude that,
although the infrastructure problems of the Web have largely been
solved, the cacophony of technologies for Web-based applications
reflects the lack of a solid model tailored for this domain.
\end{abstract}



\begin{document}

\maketitle

\setcounter{page}{1}

\begin{bottomstuff}
Author's addresses: 
B. J. Doyle, Donald Bren School of Information and Computer Sciences,
University of California, Irvine, Irvine, CA 92697; email: bdoyle@ics.uci.edu;
C. V. Lopes, Donald Bren School of Information and Computer Sciences,
University of California, Irvine, Irvine, CA 92697; email: lopes@ics.uci.edu.
\newline
\end{bottomstuff}

\maketitle

\section{THE DEMAND FOR DYNAMIC CONTENT}
Within a decade of its inception, the Web evolved from a static hypertext 
presentation medium into a standard user interface technology for a growing 
class of interactive information systems.  The initial Web implementation, 
defined by its static nature and a purposefully low barrier to entry, was 
sufficient for sharing documents but inadequate for more advanced applications.  
The early Web, despite its static limitations, demonstrated clearly enough 
the overwhelming potential and global reach of the new medium, setting in 
motion a rush of exploration into how to bring social, educational, and 
commercial applications online.  Popular interest and media hype related to 
{\em cyberspace}, a science fiction turned plausible by the arrival of the 
Internet, raised expectations for the Web far beyond its humble origins to 
those of a universal hub of online information services 
~\cite{Bell:cacm:1997}.  Up-to-date information and effective user 
interfaces, essential requirements for online service applications, could 
not be practically implemented with only static content.   
Methods for providing dynamic content on-the-fly in response to user requests 
were developed to meet the demands of Web service applications.  Conceptually, 
dynamic content extended the early Web by providing an indirection level that 
allows users to request server {\em resources}, programs that generate documents, 
as well as documents ~\cite{Berners-Lee:web:1998}.  Powered by dynamic content 
generation systems, online commerce, education, and communication services 
became widespread, and the character of the Web became increasingly 
application-centric rather than document-centric.  At the same time client-side 
interactivity was extended far beyond its original constraints, which limited 
interaction to only simple page-based navigation through hypertext links.

\subsection{The Static Web}

The World Wide Web, commonly referred to as simply the Web, was developed in 1990 
at the European Organization for Nuclear Research (CERN) as a hypertext system 
for sharing documentation between disparate computers and applications over the 
Internet ~\cite{Berners-Lee:web:1989,Berners-Lee:cacm:1994,Berners-Lee:book:2000}.  
The simple and open design of the Web attracted attention outside of CERN, 
initially mainly within the network research community. Acceptance of the Web 
accelerated after CERN released its implementation into the public domain in 
April 1993, which allowed organizations to experiment with the Web without 
concern for future licensing entanglements.

NCSA Mosaic 1.0, the first Web browser with a graphical user interface (GUI), 
was released by the National Center for Supercomputer Applications (NCSA) at the 
University of Illinois at Urbana-Champaign in September 1993. The NCSA Mosaic 
browser was simple to install and use on the X Windows, Microsoft Windows, and 
Macintosh platforms. The GUI browser was an immediate popular success that was a 
catalyst for a rush of business interests and the public to embrace the 
Internet.

The success of NCSA Mosaic and the Web led to demands for dynamic content, improved 
interfaces, and connectivity to external systems. The age of purely static Web was 
effectively over when NCSA introduced the Common Gateway Interface (CGI) in late 1993 
as a feature of the NCSA httpd 1.0 Web server. CGI allows a browser client to request 
data from a program running on a Web server rather than from a static document.  Other 
innovations such as browser fill-in forms and Server Side Includes (SSI) were 
additional early methods for making the Web more interactive and dynamic. Building on 
the early foundations provided by CERN and NCSA, the history of Web programming can be 
viewed as a continuing quest for improvements in efficiency, ease of use and reliability 
for interactive dynamic content generation.  

\subsection{Web Technology Design Pressures}
The Web architecture has evolved into an effective infrastructure for
a wide class of complex, interactive services. Several factors shaped
the mostly ad-hoc evolution of the Web. Table~\ref{tab:designpoints}
summarizes the three major factors and some of their pressures on the
evolution of the Web. This table provides a set of criteria that is used
throughout the paper for assessing and comparing dynamic Web content
development technologies.

\begin{table}
\begin{center}
\begin{tabular}{|l||l|} \hline
Factor  & Objectives \\ \hline \hline
\multirow{7}{*}{Global distribution} 
   & Reliability \\
   & Extensibility \\
   & Scalability \\
   & Performance \\
   & Portability over heterogeneity \\
   & Loose coupling \\
   & Security \\ \hline
\multirow{5}{*}{Interactivity} 
   & Dynamic page generation \\ 
   & Data validation \\
   & Handling of browser navigation anomalies \\
   & State management \\
   & Event-action \\ \hline
\multirow{5}{*}{Human and socio-economic factors} 
   & Agile development \\
   & Designer/programmer integration \\
   & Learning effort \\
   & Popularity \\ 
   & Conformance to standards and practices \\ \hline
\end{tabular}
\caption{Web technology design factors.}
\label{tab:designpoints}
\end{center}
\end{table}

{\em Global Distribution.}
The Web is a distributed system that loosely couples clients with
geographically-dispersed servers through message-passing over the
Internet.  The HTTP protocol is a specific implementation of the
general distributed systems client-server communications model. Web
applications are implemented within a distributed application layer
above HTTP, and are therefore bound to the requirements of distributed
systems.  Properties and design points for distributed operating 
systems that are important for the Web are reliability, 
extensibility, scalability, performance, heterogeneity, and security 
~\cite{Sinha:book:1997}.

{\em Interactivity.}
Working around the user interface interactivity limitations of 
the Web architecture is a central theme of Web programming.
Constraints of the HTTP protocol, which impose a stateless, strictly
pull-only architecture on Web applications, mandate page-centered
interactivity, limiting client processing to page rendering, 
form data collection, and navigation to subsequent pages.  Any 
interaction between a client and a server results in a complete 
page to be downloaded in response, even if the new page differs only 
slightly from the previous version. The level of interactivity is 
similar to the archaic block-mode interfaces that connect dumb 
terminals to mainframe computers. The need to address the limitations 
of HTTP, which are justifiable for scalability and simplicity reasons, 
has provided much of the impetus for the development of technologies 
that have made the Web progressively more reliable, interactive, 
and dynamic.

{\em Human Factors.}
Additional requirements for Web development technologies are related
to human factors that reflect the cultural and social environment in
which the applications are created. 

\subsection{Objective}
The transition to the modern dynamic Web and beyond has generally been neither 
organized nor orderly, rather it was accomplished through independent innovation by 
many individuals and groups; sometimes collaboratively, sometimes competitively, and 
often in isolation.  Web developers are pragmatic when evaluating new technologies: 
those that work well within the constraints of the applications are accepted; others are 
ignored or become obsolete, not explicitly retired, but instead gradually falling 
out of use.  As a result, the Web development landscape is cluttered
with a variety of technologies and tools, and littered with obsolete or 
failed options that can trap newcomers.

The objective of this paper is to derive order from the chaos by
describing the foundations of the Web and classifying the related
technologies and programming techniques that are used to create
dynamic Web applications. Although server-side dynamic content
generation is an essential part of almost all Web applications, it is
too enmeshed with other concerns to be adequately explored in
isolation.  Therefore, the remaining sections of this paper describe
the relevant technologies and techniques separately and as they
evolved in relation to each another.  In Section \ref{sec:support},
the Web's architecture is reviewed, since it is the foundation for the
technologies that support dynamic content. The progression of
developments that made the Web increasingly dynamic and reliable is
reviewed to survey important technologies and the context in which
they were created. Section \ref{sec:enterprise} surveys technologies
specifically created for large enterprise software systems. A summary
and classification of the technologies is presented in section
\ref{sec:classification}, providing a roadmap for technology
selection. Section \ref{sec:programming} explores software engineering
methods and tools that have been carried forward from traditional
programming into the realm of the Web, and assesses their fitness and
level of maturity for developing Web applications. Finally, section
\ref{sec:conclusions} draws the conclusions of our study.


\section{SUPPORT FOR DYNAMIC CONTENT}
\label{sec:support}

Technologies are best understood in light of the context that motivated their creation.  
The evolution of the Web has been primarily motivated by a search for abstractions that 
improve its usability for end users, through richer user interfaces, and for developers 
with more powerful programming methods. This section recounts the road to 
dynamic content as a sequential narrative, starting with a brief review of the 
architectural foundations of the Web and design points for Web applications.  The 
remaining sub-sections describe the progression of developments that shaped the dynamic 
Web, from the simple initial gateway methods that enabled dynamism, to the recent 
frameworks that aim to simplify development while improving reliability and 
functionality.   

\subsection{An Architecture for Distributed Hypermedia Applications}
The architectural foundation of the Web is the request-response 
cycle realized by the Hypertext Transfer Protocol (HTTP) 
~\cite{Fielding:web:1999} and Hypertext Markup Language (HTML and XHTML) 
~\cite{Raggett:web:1999,Altheim:web:2001} standards.  The 
Representational State Transfer (REST) architectural style provides a 
model architecture for the Web that was used to rationalize the definition
of the HTTP 1.1 recommendation ~\cite{Fielding:acmtoit:2002}.  
Modularity, extensibility, and inversion of control are characteristics 
inherent in the Web that have allowed incorporation of features supporting 
dynamic content. Inversion of control is implemented on both clients 
and servers by various plug-in, content handling, and filtering interfaces 
that allow custom components to be invoked in response to events.  
The following sections review the operations of Web servers and 
browsers highlighting the extension points that are leveraged to provide 
interactive and dynamic content for Web applications.  

\subsubsection{Web Servers}
\begin{figure}
\centerline{\includegraphics[height=5cm]{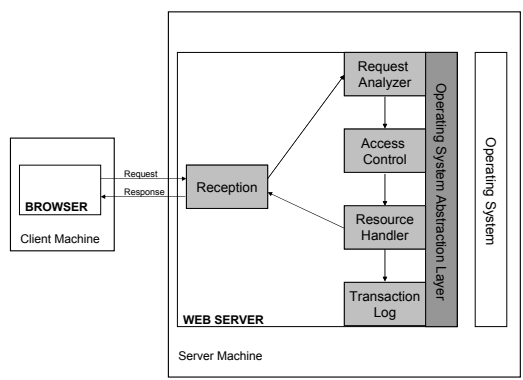}}
\caption{A reference architecture for Web servers adapted from the architecture 
of Hassan and Holt [2000].}
\label{fig:WebServerRefArch}
\end{figure}

Web servers implement the server-side duties of HTTP, the application-layer protocol that 
governs message-passing between Web clients and servers. Figure~\ref{fig:WebServerRefArch} 
is adapted from a reference architecture for Web servers provided by 
~\citeN{conf:wcre:Hassan:2000}. The most common Web server implementations are the Apache 
HTTP server ~\cite{Apache:web:2004}, available for most operating systems, and the 
Internet Information Service (IIS) ~\cite{Microsoft:web:2005}, available only for 
Microsoft Windows operating systems.  Request and response messages share a common format 
that includes a start line, message headers, and optionally, a message body and message 
trailers.  Request messages specify a request method, most commonly {\tt GET} or 
{\tt POST}, and a Universal Resource Identifier (URI) for a requested resource.  
Resources are a key abstraction for the Web, uniformly identifying documents, 
services, collections of other resources, and other types of information sources using 
a single naming scheme. Response messages include a status line and a  
representation of a resource.  The protocol supports transmission of any content type 
that can be represented as a sequence of bytes with associated metadata.  Responses 
are interpreted by client browsers.

\subsubsection{Web Browsers}
\begin{figure}
\centerline{\includegraphics[height=5cm]{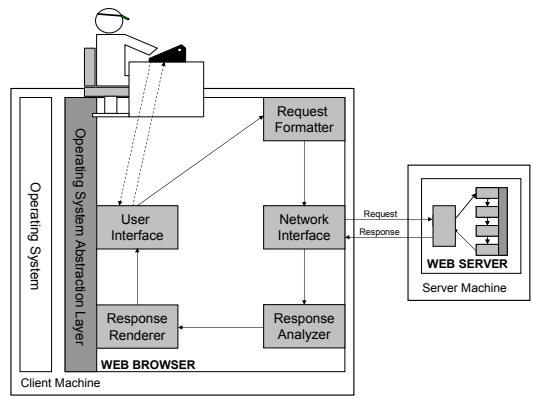}}
\caption{An architecure that illustrates the operations of Web browsers.}
\label{fig:WebBrowserRefArch}
\end{figure}

Web browsers process user interface commands, format and send request 
messages to Web servers, wait for and interpret server response messages, 
and render content within the browser's display window area. 
Figure~\ref{fig:WebBrowserRefArch} is a simplified architecture diagram 
that illustrates the operations of Web browsers. HTML and XHTML 
~\cite{Altheim:web:2001} are the most common content types on the Web. 
Browser extensibility features allow many other content types to be displayed 
by deferring their rendering to registered plug-in components (helper 
applications) that handle the content. The pervasiveness of HTML makes it 
a friendly target for dynamic content generation systems. 

\subsubsection{HTML}
HTML documents contain text interspersed with formatting elements.
{\em Cascading Style Sheets (CSS)} allow formatting elements to be 
separated into reusable style sheets ~\cite{Bos:web:2004}. Uneven CSS
implementations in browsers and ingrained practices resulted in an
extended acceptance period for style sheets. Promotion of Web
standards and improved browser implementations of the standards have
recently resulted in steadily increasing use of CSS to separate
presentation attributes from content ~\cite{Zeldman:book:2003}.

Client-side scripts can be embedded in HTML documents to add interactivity 
to Web pages and further process a document before it is rendered by a browser. 
The {\em Document Object Model (DOM)} ~\cite{W3C:web:2004} interface allows 
embedded scripts to modify the structure, content, and presentation of 
documents. The combination of HTML, client-side scripting, and DOM is 
informally known as {\em Dynamic HTML (DHTML)}. The most popular 
client-side scripting language is JavaScript. It is also possible 
to reference Java applets, ActiveX controls, Macromedia Flash presentations,
and other kinds of precompiled programs within HTML documents, but the 
approach has compatibility, latency, and security issues that limit its effectiveness 
~\cite{Bellovin:web:2000}. In spite of these concerns, ActiveX and Macromedia 
Flash are still widely used by web designers to provide a more 
graphically-intensive user experience than would otherwise be practically
achievable on the Web. While the promising W3C Scalable Vector Graphics 
(SVG) ~\cite{W3C:web:2005} and Synchronized Media Integration Language 
(SMIL) ~\cite{W3C:web:2005:b} standards provide XML-based alternatives to 
Macromedia Flash for multimedia presentations, they are not yet pervasively 
used.

\subsubsection{XML}
The {\em Extensible Markup Language (XML)} is a widely accepted markup 
language that simplifies the transmission of structured data between 
applications ~\cite{Yergeau:web:2004}. XML is a meta-language for creating 
collections of custom elements, in contrast to HTML, which provides a fixed 
set of elements. The {\em Extensible Stylesheet Language (XSL)} family 
includes an XML-based element matching language for XSL Transformations 
(XSLT) that is used to programmatically transform XML documents into other 
formats ~\cite{Clark:web:1999}. 

XML has been extremely successful standard since the original
recommendation was released in December, 1997. XML provides the base
syntax for the XHTML and CSS standards that normalize and will
eventually replace HTML as the dominant presentation technology for
the Web. Configurations for Web applications and services are now commonly
maintained within XML files. The extensive availability of XML parsers
makes it more convenient for programmers to work with XML files rather
than develop parsers for proprietary file formats.  Web services standards 
including SOAP ~\cite{W3C:web:2004:b} and XML-RPC ~\cite{Winer:web:1999} 
leverage XML for configuration and as a request and response message format 
for remote procedure calls over HTTP. 

\subsection{Initial Dynamism}
The earliest technical elements that allowed for interactive and dynamic
content were HTML forms, the HTTP {\tt POST} request method, and the
Common Gateway Interface (CGI). HTML forms are used to collect user
input data that is submitted to a forms processor on a Web server in a
{\tt GET} or {\tt POST} message. By 1993, the availability of CGI
completed the forms processing path by providing a means by which Web
servers could process and respond to submitted data. CGI is functional 
but not scalable; as its limitations became clear other solutions were 
developed that were more efficient but more complicated. This section 
reviews the first wave of technologies for the dynamic Web including 
CGI, its server-side successors, and client-side extension interfaces.

\subsubsection{Forms}
The HTML forms capability naturally extends the Web's document metaphor 
by allowing user input to be entered on Web pages. A form is a section of a 
document that contains named user interface controls such as text boxes, 
check boxes, radio buttons, list boxes, and buttons ~\cite{Raggett:book:1997}. 
The definition of a form specifies a request method ({\tt GET} or {\tt POST}) 
and a URI for a server-side forms processor. When a form is submitted, the 
browser formats a request message containing the form data as a sequence of 
name-value pairs. For {\tt GET} messages, the form data set is appended to 
the action URI as query parameters. When {\tt POST} is used, the form data 
is sent in the message body. 

The forms capability of HTML is relied on by many Web applications. The  
forms interface is simple, cross-platform, supports light data validation, 
and allows pages to be event-driven. The event model of a form is implicit
in the URI references associated with submit buttons. The loosely coupled 
interaction between forms and their processors can be a source of reliability 
problems since there is no static guarantee that the data types of  
submitted data elements conform to the expectations of form processor.

\subsubsection{CGI}
\begin{figure}
\centerline{\includegraphics[height=5cm]{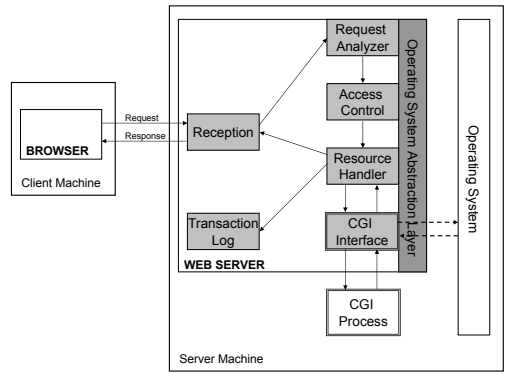}}
\caption{The CGI architecture.}
\label{fig:cgiarch}
\end{figure}

CGI was the first widely available means for integrating Web servers
with external systems, initially provided as a method to process data
submitted from HTML forms ~\cite{NCSA:web:1993}.  CGI allows
server-side programs to be invoked in response to HTTP requests. A 
Web server creates a new process for each CGI request.  
Figure~\ref{fig:cgiarch} shows the archictecture of CGI.  CGI programs 
can be written in any programming language that supports environment 
variables and the standard input and output streams.  The earliest CGI
programs were written in C, but the deployment ease and portability of
interpreted scripting languages such as tcl, Perl, and Python has made
them the languages of choice for CGI ~\cite{Ousterhout:ieeec:1998}.  
Perl is the most popular language for CGI scripting.  User input and 
metadata about requests is passed into CGI programs through
environment variables and within the standard input stream,
respectively.  The output written by a CGI program to its standard
output stream is sent to the client within an HTTP response message.
The example Perl script in Figure~\ref{fig:cgi} reads an environment
variable to determine the request method ({\tt GET} or {\tt POST}) and
displays the data that was submitted from a form.

\begin{figure}
\centering
\begin{tabbing}
{\#}! /usr/local/bin/perl \\
{\#} Display the form data set sent in a GET or POST request. \\ 
\\
{\bf print} "Content-type: text/html$\backslash $n$\backslash $n"; \\
{\bf print} "$<$html$><$head$><$title$>$Form Data$<$/title$><$/head$>\backslash $n"; \\
{\bf print} "$<$body bgcolor=$\backslash $"{\#}FFFFFF$\backslash $"$\backslash $n$>$" \\
\\
{\bf if} ({\$}ENV{\{}'REQUEST{\_}METHOD'{\}} eq 'POST') {\{} \\ 
    \hspace*{2em}\= \+ read (STDIN, {\$}buffer, {\$}ENV{\{}'CONTENT{\_}LENGTH'{\}}); \\ 
    @pairs = split(/{\&}/, {\$}buffer); \-\\ 
{\}} {\bf elsif} ({\$}ENV{\{}'REQUEST{\_}METHOD'{\}} eq 'GET') {\{} \+\\ 
    @pairs = split(/{\&}/, {\$}ENV{\{}'QUERY{\_}STRING'{\}}); \-\\ 
{\}} {\bf else} {\{} \+\\ 
    {\bf print} "$<$p$>${\$}ENV{\{}'REQUEST{\_}METHOD'{\}} message received$<$/p$>$"; \-\\ 
{\}}
\\
{\bf foreach} {\$}pair (@pairs) {\{} \+\\ 
    ({\$}name, {\$}value) = split(/=/, {\$}pair); \\
    {\$}value =$\sim $ tr/+/ /; \\
    {\$}value =$\sim $ s/{\%}([a-fA-F0-9][a-fA-F0-9])/pack("C", hex({\$}1))/eg; \\
    {\$}name =$\sim $ tr/+/ /; \\
    {\$}name =$\sim $ s/{\%}([a-fA-F0-9][a-fA-F0-9])/pack("C", hex({\$}1))/eg; \\
    {\bf print} "$<$p$>$Field {\$}name has the value {\$}value$<$/p$>\backslash $n"; \\ 
    {\$}FORM{\{}{\$}name{\}} = {\$}value; \-\\ 
{\}} 
{\bf print} "$<$/body$><$/html$>\backslash $n"; 
\end{tabbing}
\caption{An example of a Perl CGI script.}
\label{fig:cgi}
\end{figure}

CGI was the first widely supported technology for dynamic content and is 
still supported out-of-the-box by most Web servers.  In tandem with 
scripting languages, CGI is a platform-independent solution with a 
simple, well-known interface.  The disadvantages are related to scalability 
and usability concerns.  CGI is not highly scalable because a new process 
must be created for each request.  For busy Web sites serving thousands of 
concurrent users, the CPU and memory usage required to constantly create and 
destroy processes severely limits the number of concurrent requests that 
can be handled.  The use of scripting languages further strains a Web server's 
capacity due to the need to start an interpreter for each request. 

The usability problems of CGI stem from the limitations of its thin 
abstraction over the HTTP protocol.  Programmers must understand the workings 
of HTTP, down to the level of formatting details of resource identifiers and 
messages, to be able to create CGI scripts.  No page computation model is 
provided; the programmer is responsible for generation of the response by printing 
HTML to the standard output stream.  Role separation between designers and 
programmers is diminished due to the fact that the presentation attributes 
of pages are embedded with print statements in programs.  Web page authoring 
tools such as FrontPage or Dreamweaver can not be used since the presentation 
HTML is embedded within a program's logic. 

Other responsibilities including state management, security, validation, data 
access, and event handling are completely delegated to 
programmers.  A spate of fragile, idiosyncratic Web application implementations 
were the result of the lack of structure allowed by CGI.  The introduction of 
software engineering discipline in the form of coding guidelines, scripting libraries, 
and frameworks has improved the situation to some extent ~\cite{Stein:book:1998}.

Despite its limitations, CGI is not obsolete. It natively exists within most 
Web servers, in contrast to other dynamic content solutions that require additional 
component installation.  The out-of-the-box, universal availability of CGI makes 
it a possible target for small to medium-sized applications with low-volume 
expectations.  ~\citeN{conf:scc:Wu:2000} found CGI to be inefficient in handling 
concurrent client requests and therefore suitable only for low-traffic applications 
based on benchmark comparisons to other options for generating dynamic
content.  The technology is still in use mainly due to the increasing popularity 
of scripting languages, which can provide a straightforward, portable alternative 
to Java.  

\subsubsection{Scalable CGI Implementations}
FastCGI (Open Market), mod{\_}perl combined with the Apache::Registry 
module (Apache), and PerlEx (ActiveState) are examples of Web server 
extensions that improve the scalability of CGI. FastCGI is a CGI 
implementation that maintains a pool of persistent processes that are reused 
for multiple requests to reduce process creation overhead ~\cite{Brown:web:1996}.
Figure~\ref{fig:cgiarch} shows the architecture of scalable CGI implementations. 
mod{\_}perl is an Apache extension that embeds a Perl interpreter within the 
Web server that allows Perl scripts to access the Apache C language API. 
Apache::Registry is a Perl library that supports CGI under mod{\_}perl. 
The combination of mod{\_}perl and Apache::Registry improves performance by 
avoiding the overhead of starting and stopping a Perl interpreter for each request. 
An Apache Web server can also be extended to support corresponding capabilities 
for other scripting languages including Python (mod\_snake, mod\_python), 
tcl (mod\_tcl), and Ruby (mod\_ruby with eRuby).  PerlEx provides similar 
capabilities for Microsoft IIS by maintaining a pool of interpreters that 
is managed by a Web server extension module ~\cite{ActiveState:manual:2003}.

\begin{figure}
\centerline{\includegraphics[height=5cm]{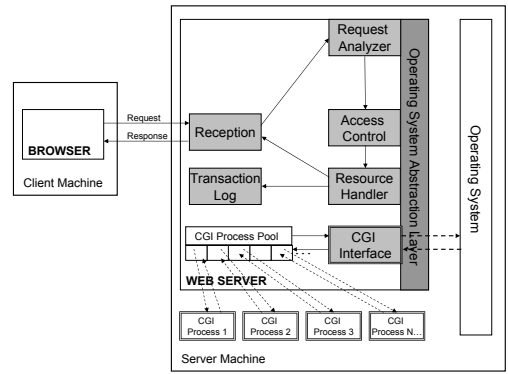}}
\caption{The scalable CGI architecture.}
\label{fig:scalablecgiArch}
\end{figure}


~\citeN{conf:sane:Gousios:2002} compared the performance of FastCGI, 
mod{\_}perl, PHP, and Java servlets under Apache on Linux using 
a minimal commodity hardware configuration (a single Pentium III 733 MHz 
processor with 384 MB of memory).  Their results showed that FastCGI was 
the best performing and most reliable option on the benchmark hardware. 
Java servlets also performed steadily, even though the authors conceded 
that the benchmark conditions were not realistic for the technology,
which is more appropriately matched to enterprise-level hardware 
supporting multiple processors and larger amounts of memory. 
~\citeN{Apte:cc:2003} compared the performance of a similar technology
group (CGI, FastCGI, Java servlets, and JSP) on a dual-processor Solaris
system (2 360 MHz Sun Ultra processors with 512 MB of memory) with 
similar results that showed FastCGI to be the best performer. However, the
authors also concluded that factors other than performance, including 
development time, support availability, ease of integration, and 
deployment convenience, are also important concerns for Web 
development groups.

\subsubsection{Web Server Extension Interfaces}
The initial reaction from Web server providers to CGI performance 
issues was to create proprietary APIs with similar capabilities 
~\cite{Reilly:book:2000}. All of the major Web servers have APIs that 
can be used to introduce new functionality into a Web server through 
extension modules. The most well-known examples are NSAPI, originally 
created for Netscape Web servers ~\cite{Sun:manual:2003}; Microsoft 
ISAPI ~\cite{Microsoft:web:2005:b}; and the Apache module API 
~\cite{asf:manual:2003}. Extension modules are usually required to be 
written in C or C++ and compiled into dynamic link libraries that are 
linked into the Web server at runtime.  Extensions can run extremely 
fast since they are compiled code that runs in the Web server address space.

\begin{figure}
\centerline{\includegraphics[height=5cm]{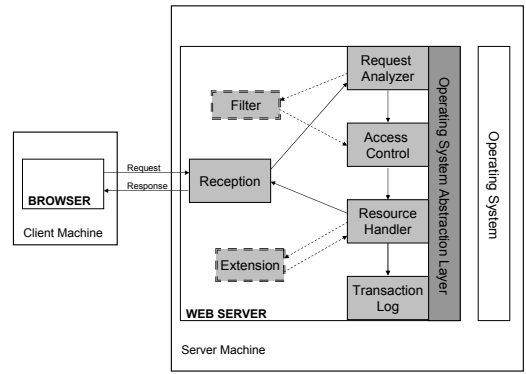}}
\caption{The reference architecture for Web servers modified to show filters
  and extensions.}
\label{fig:webServerExtension}
\end{figure}

ISAPI and the Apache interface are representative of Web server APIs in 
general. ISAPI supports the development of {\em extensions} and 
{\em filters} that modify the behavior of IIS. 
Figure~\ref{fig:webServerExtension} shows the placement of filters 
and extensions within the reference architecture. The corresponding 
Apache API constructs are {\em modules}, {\em handlers}, and 
{\em filters} ~\cite{Thau:cn:1996}. ISAPI extensions behave like 
CGI scripts; extensions are invoked directly by clients or through 
URI mapping, and are responsible for handling requests and 
creating responses. On IIS servers, a well-known example is the mapping of 
{\tt .asp} files to {\tt asp.dll}, the Active Server Pages 
interpreter. A corresponding example for Apache is the association of 
{\tt .php} files to the mod{\_}php extension module. ISAPI filters 
perform additional behaviors in addition to the default behaviors, and 
can be used to implement custom logging, authentication, mapping, and 
retrieval features. The Apache API also supports filters as a modular way to 
manipulate request or response data streams.

Web server APIs were originally designed as scalable replacements for CGI, 
but they are rarely directly used to build Web applications. The APIs are 
complex, non-portable, and require advanced programming knowledge, so 
extension modules are difficult to build, test, and maintain. Reliability 
can be compromised due to the tight integration of extensions into Web 
servers; a single flaw in an extension module can crash a Web server. The 
cost of developing extensions is easier to justify for widely reusable 
features than for those supporting only a single application.  In spite 
of their weaknesses, Web server APIs are an important building 
block for dynamic content generation systems. In fact, for performance 
reasons most server-side technologies that support dynamic content are 
based on Web server extension modules. 

\subsubsection{Browser Extension Interfaces}
One of the first browser extension interfaces was the Common Client 
Interface (CCI) for NCSA Mosaic ~\cite{NCSA:web:1995}. CCI was an API that allowed 
external applications to communicate with running browser instances by requesting 
a URL to be displayed. CCI is obsolete but influenced the browser extension 
technologies that followed. 

\begin{figure}
\centerline{\includegraphics[height=5cm]{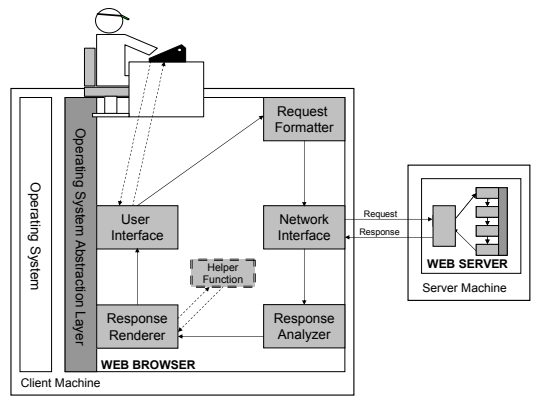}}
\caption{Browsers are extended through plug-in and scripting interfaces that
  influence the rendering and handling of the user interface.}
\label{fig:browserExtension}
\end{figure}

During the browser wars of the mid-1990s all of the major browser 
providers created proprietary APIs to differentiate their products. The 
Netscape Plug-In API and Microsoft ActiveX are examples of 
browser-specific APIs. For systems requiring dynamic interfaces, the key 
features of browser-specific APIs are support for plug-in components and 
access to internal browser methods and properties related to the 
presentation of content. Security that prevents downloaded components from 
performing undesirable actions is a key requirement for browser extensions. 
ActiveX makes use of ``Authenticode'', a code-signing scheme that verifies 
that downloaded binary components are pristine as offered by a certified 
provider prior to their execution by a browser. Figure~\ref{fig:browserExtension} 
illustrates the place of extensions within the architecture of browsers.

\noindent
{\bf Applets.}
Java applets represent an extension approach that is not browser-specific 
since it leverages the portable Java byte code format. Applets are Java 
class files that are downloaded and interpreted by a Java Virtual Machine 
(JVM) running within the  browser address space. The JVM executes applets 
only after verifying that their code is safe, meaning that it has not been 
tampered with and contains no instructions that violate the client's 
security policy. Applets can also be digitally signed and verified to 
provide an additional level of security. Java applets initially suffered 
from poor perceived performance due to extended download times and 
slow interpretation; therefore the technology has been relegated to a 
secondary role, even though performance has since been vastly improved by 
the introduction of just in time (JIT) compilation to native code. The 
pervasive Macromedia Flash animation player plug-in provides 
an alternative to Java applets that is now commonly used to embed marketing
presentations in Web pages.

\noindent
{\bf Client-side Scripting.}
Interpreters for lightweight scripting languages such as JavaScript
and VBScript were available for most browsers by the end of
1996. Client-side scripting languages interfaces are more accessible
than browser extension APIs, since they remove the need to know an
entire API to add pieces of useful functionality to an HTML
document. Client-side scripts are slightly less efficient than
plug-ins, but the advantages of easy access to browser properties and
methods outweigh the performance penalty. Initially, each browser
creator implemented a proprietary scripting language and API that was
incompatible by design with other browsers. Scripting language
standards, including ECMAScript and DOM, have improved the situation
to the point that cross-browser scripting is possible.

\noindent
{\bf Rich Internet Applications.}
The applet concept has recently been revived under the guise 
of {\em rich Internet applications (RIA)}. The 
objective of the RIA concept is to break away from the page-centered 
constraints imposed by the Web to support a more interactive user 
experience.  RIA solutions that attempt to introduce a new 
plug-in architecture on to the Web (Konfabulator, Curl, and Sash 
Weblications, to name a few) have attracted attention, but eventually lose 
momentum due to the requirement to download and install a plug-in. Laszlo 
and Macromedia Flex are RIA environments that are attempting to exploit the 
large installed base of Flash users to provide improved interactivity 
without requiring plug-in installation. Laszlo applications are described 
using an XML application description language. The Laszlo Presentation 
Server is a Java servlet that dynamically composes and serves Flash 
interfaces in response to requests for Laszlo applications. RIA solutions 
can improve the responsiveness and presentation quality of Web user 
interfaces, but have not reached the mainstream of development. More 
experience with the technologies is needed to assess their compatibility 
with the existing Web infrastructure before widespread adoption of RIA will 
occur.  

\noindent
{\bf Expanded Use of Dynamic HTML.}
~\citeN{Garrett:web:2005} coined the term {\em Ajax} for using the 
combination of XHTML, CSS, DOM, XML, XSLT, JavaScript, and XMLHttpRequest, 
a JavaScript API for accessing Web services, to deliver RIA completely within 
the existing Web infrastructure.  The production application of Ajax within 
several popular, high-volume web sites including Gmail, Google Suggest, 
Google Maps, and the Amazon.com A9 search engine provides evidence that the 
combination can be effective and scalable.  The disadvantages of Ajax lie 
in browser compatibility issues and the non-straightforward JavaScript 
coding that can be required to implement simple functionality.

\subsection{Interpreted Template-Based Scripting}
\begin{figure}
\centerline{\includegraphics[height=5cm]{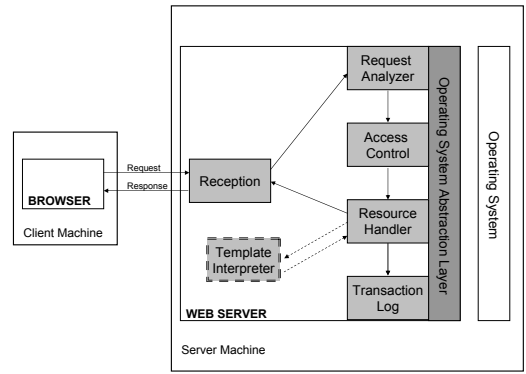}}
\caption{The Web server architecture is extended with a template-based
  scripting interpreter to implement server-side templates.}
\label{fig:webServerTemplates}
\end{figure}

The use of templates is a common characteristic of pattern-based program 
construction systems. A {\em template} is a model for a set of documents, 
composed of fixed and variable parts, that is contextually expanded to 
yield conforming document instances. The variable parts of a template 
encapsulate programming logic such that each variable part exposes a 
program function that specifies how it is expanded in context. 
Predominately-fixed templates are easier to construct and comprehend than 
more variable templates since less programming is required and the defined 
structure is largely static. Many Web sites consist of generally fixed HTML 
pages with small amounts of variable content. In contrast to template 
processing, CGI processing is more suitable for applications with mostly 
variable content, which are not as common on the Web. The template model 
better supports the common Web application pattern by embedding logic within 
fixed presentation markup. Templates reduce the programming skill needed 
to create dynamic content. Role separation is well-supported since the addition 
of logic can be delegated to programmers. 
Figure~\ref{fig:webServerTemplates} illustrates that template-based
scripting interpreters are implemented as Web server extensions.

\subsubsection{Server-side Includes}
The \textit{Server Side Includes (SSI)} capability was introduced by
NCSA in 1993 as method for embedding trivial computations within HTML
pages to be processed directly by the Web server, instead of by a
separate process as with CGI.  SSI templates contain fixed HTML text 
and variable areas with commands that execute before a response is sent 
to a user. The technology is {\em tag-centered}, in that dynamic behavior 
is specified by special HTML tags, formatted comments that denote command 
keywords and parameters.

The initial SSI implementation supported the {\tt include} command for text file 
inclusion, the {\tt echo} command for variable output, and the {\tt exec} command 
for running external programs. Conditional processing and other features were 
independently added to SSI by Web server providers. On Apache Web servers, 
SSI plus several extensions is known as {\em Extended SSI (XSSI)}.  
Figure~\ref{fig:xssi} is an example Apache XSSI template with conditional processing 
commands ({\tt if}, {\tt elif}, and {\tt endif} commands) and core SSI commands 
including {\tt config}, {\tt include}, and {\tt echo}. The example shows two typical 
uses of SSI which are to include common elements on a group of pages, and content 
adaptation based on client attributes.

\begin{figure}
\centering
\begin{tabbing}
$<$html$>$ \\
$<$head$>$ \\
   \hspace*{2em}\= \+ $<$meta http-equiv="Content-Language" content="en-us"$>$ \\
   $<$title$>$SSI Example$<$/title$>$ \\
   {\em $<$!--{\tt {\#}set var=}"VAR{\_}css" value="msie" --$>$} \\
   {\em $<$!--{\tt {\#}if expr=}"({\$}HTTP{\_}USER{\_}AGENT=/Mozilla/)} \\
   \hspace*{2em}{\em {\tt{\&}{\&}} ({\$}HTTP{\_}USER{\_}AGENT !=/compatible/)" --$>$} \\
   {\em $<$!--{\tt {\#}set var=}"VAR{\_}css" value="nav" --$>$} \\
   {\em $<$!--{\tt {\#}elif expr=}"({\$}HTTP{\_}USER{\_}AGENT=/Opera/)" --$>$}  \\
   {\em $<$!--{\tt {\#}set var=}"VAR{\_}css" value="opera" --$>$}  \\
   {\em $<$!--{\tt {\#}endif} --$>$} \\
   $<$LINK REL="stylesheet" type="text/css" \\
   href="/css\{{\em /$<$!--{\tt {\#}echo var=}'VAR{\_}css' --$>$.css"$>$} \-\\
$</$head$>$ \\
$<$body$>$ \+\\
   {\em $<$!--{\tt {\#}include virtual=}"pageheader.shtml" --$>$ } \\
   $<$p$>$This is an example SSI page.$<$/p$>$ \\
   $<$p$>$Document name: {\em $<$!--{\tt {\#}echo var=}"DOCUMENT{\_}NAME" --$>$} $<$/p$>$ \\
   $<$p$>$Server local time:{\em $<$!--{\tt {\#}config timefmt=}"{\%}I:{\%}M {\%}p {\%}Z" --$>$ }  \\
   {\em $<$!--{\tt {\#}echo var}="DATE{\_}LOCAL" --$>$} \\ 
   $</$p$>$ \\
   $<$p$>$Browser type: {\em $<$!--{\tt {\#}echo var=}"HTTP{\_}USER{\_}AGENT"--$>$} $<$/p$>$ \\ 
   {\em $<$!--{\tt {\#}include virtual=}"pagefooter.shtml" --$>$ } \\
   $<$p$>$Last updated:{\em  $<$!--{\tt {\#}config timefmt=}"{\%}c" --$>$  \\
   $<$!--{\tt {\#}echo var=}"LAST{\_}MODIFIED" --$>$} $<$/p$>$ \-\\
$</$body$>$ \\
$</$html$>$ 
\end{tabbing}
\caption{An example of an XSSI document.}
\label{fig:xssi}
\end{figure}

While infrequently used, SSI is not completely obsolete and the core set of 
commands is supported by most Web servers. Role separation is better 
supported by SSI than by CGI since the fixed and variable areas of pages can 
be built independently. Web designers can work strictly with HTML to design 
the appearance of a page, adding {\tt exec} commands to retrieve information where 
dynamic content is required. Development of logic invoked by {\tt exec} commands can 
be delegated to programmers. 

The disadvantages of SSI are related to scalability and usability concerns. 
SSI is no more scalable than CGI, and can be worse for documents with multiple 
{\tt exec} commands. Each {\tt exec} command leads to a process creation, adding 
appreciably to the Web server processing load even relative to CGI. The usability 
concerns are due to the primitive syntax and CGI-based process model. The syntax is 
error-prone, difficult to internalize, but does not not support complex processing 
other than through CGI via {\tt exec} commands. The dependence on {\tt exec} for 
non-trivial processing defeats role separation. For larger 
applications, the inevitable reliance on {\tt exec} leads to barely maintainable, 
multi-language tangles of templates, scripts, and programs. While SSI is rarely 
appropriate for new applications, it provides an accessible, low cost alternative
for non-critical, low-traffic Web applications

\subsubsection{ColdFusion}
ColdFusion was introduced in 1995 by Allaire Corporation as an ``easy to program'' 
technology for creating dynamic content. The defining feature of the technology is the 
template-based ColdFusion Markup Language (CFML), originally interpreted but now 
JIT-compiled into servlets. CFML advanced the templating model of SSI by providing a 
large set of commands, originally focused on database processing but ultimately also 
supporting other functions including file management, flow control, and forms processing, 
that is designed to be comprehensive for Web applications. The command syntax is simple, 
in that ColdFusion commands are recognizable as tags with names starting with {\tt cf} 
with variable references enclosed in {\tt {\#}} characters.  Figure~\ref{fig:cfml} shows 
an example of a simple ColdFusion template that displays dynamic content retrieved 
from a database. 


\begin{figure}
\centering
\begin{tabbing}
{\em $<${\tt cfquery name=}"AuthorResult" datasource="bookdb"$>$} \\
      \hspace*{2em}\= \+ {\em SELECT authorName FROM authors} \-\\
{\em $</${\tt cfquery}$>$} \\
$<$html$>$ \+\\
   $<$head$>$ \\
      \hspace*{2em}\= \+ $<$title$>$ColdFusion Example Author Listing$<$/TITLE$>$ \-\\
   $</$head$>$ \\
   $<$body$>$ \+\\
      $<$h1$>$Author List$<$/h1$>$ \\
      {\em $<${\tt cfoutput query=}"AuthorResult"$>${\#}authorName{\#}$<$BR$><$/{\tt cfoutput}$>$} \-\\
   $</$body$>$ \-\\
$</$html$>$ \\
\end{tabbing}
\caption{An example of a ColdFusion document.}
\label{fig:cfml}
\end{figure}

ColdFusion is comparable to SSI, sharing many of its advantages and disadvantages. 
Templates are portable since interpreters are available for several Web servers 
including Apache and IIS. Role separation between Web designers and programmers is 
possible to a similar extent as with SSI. Database access is optimized compared 
to CGI and SSI. The syntax is more expressive than that of SSI, but still not 
well-suited for expressing complex logic. The size of the language, in terms of 
the sheer number of options, reduces its usability.

\subsubsection{Server-Side JavaScript}
\begin{figure}
\centerline{\includegraphics[height=5cm]{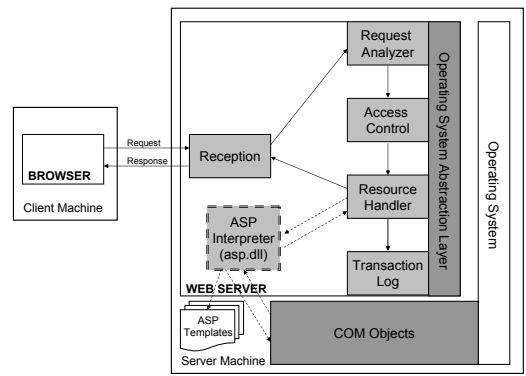}}
\caption{The architecture of ASP.  {\tt asp.dll} is the scripting language interpreter
   that provides access to COM objects for interoperability.}
\label{fig:classicASP}
\end{figure}

The next widely-used technology for server-side scripting was 
{\em script-centered templating} as exemplified by the introduction 
of {\em Server-Side JavaScript (SSJS)} and the Microsoft 
{\em Active Server Pages (ASP)} environment in 1996.  
Figure~\ref{fig:classicASP} shows the place of script-centered templating
within the reference Web server architecture.  In script-centered 
templating, blocks of logic are embedded in HTML pages. Script-centered 
templating quickly gained popularity among programmers as a more natural 
way to produce dynamic content. SSJS preceded and influenced ASP but 
did not catch on with developers and is obsolete. SSJS pages were 
constructed from HTML and JavaScript code contained within 
{\tt $<$SERVER$>$} tags. SSJS pages were compiled into byte-code 
representations that were interpreted by an interpreter on the Web 
server. The SSJS development and runtime environment provided 
{\tt server}, {\tt application},{\tt database}, {\tt client}, and 
{\tt request} objects that supported state management. Output functions 
including {\tt write} and {\tt writeln} were provided to generate 
dynamic content into HTTP responses from JavaScript code blocks. 
Figure~\ref{fig:ssjs} shows an example of a SSJS script.

\begin{figure}
\centering
\begin{tabbing}
$<$html$>$ \\
   \hspace*{2em}\= \+ $<$head$> \quad <$title$>$Server-Side JavaScript Example Author Listing$<$/title$> \quad <$/head$>$ \\
   $<$body$>$ \\
      \hspace*{2em}\= \+ $<$h1$>$Author List$<$/h1$>$ \\
      {\em $<$server$>$ } \\
         \hspace*{2em}\= \+ {\em {\tt if} (!database.connected()){\{}} \\
             \hspace*{2em}\= \+ {\em database.connect("ODBC","bookdb","admin","","");} \-\\
         {\em {\}} }  \\
         {\em {\tt if }(database.connected()) {\{} } \+\\
            {\em qs {\tt =} "SELECT au{\_}id, au{\_}fname, au{\_}lname FROM authors";} \\
            {\em results {\tt = } database.cursor(qs);} \\
            {\em write("$<$table border=2 cellpadding=2 cellspacing=2$>$" + } \\
               \hspace*{2em} {\em "$<$tr$><$th$>$ID$<$/th$><$th$>$First Name$<$/th$><$th$>$Last Name $<$/th$><$/tr$>\backslash $n"); } \\
            {\em {\tt while}(results.next()) {\{} } \\
               \hspace*{2em} \= \+ {\em write("$<$tr$><$td$>$" + results.au{\_}id + "$<$/td$>$ + "$<$td$>$" + } \\
                  \hspace*{2em} {\em results.au{\_}fname + "$<$/td$>$" + } \\
                  \hspace*{2em} {\em "$<$td$>$" + results.au{\_}lname + "$<$/td$><$/tr$>\backslash $n"); } \-\\ 
            {\em {\}}} \\
            {\em results.close(); write("$<$/table$>\backslash $n"); } \-\\
         {\em {\}}} \\
         {\em {\tt else} {\{}} \+\\
            {\em write("$<$p$>$Database connection failed");} \-\\
         {\em {\}}} \-\\ 
      {\em $</$server$>$ } \-\\
   $</$body$>$ \-\\
$</$html$>$ 
\end{tabbing}
\caption{An example of a Server-Side JavaScript template.}
\label{fig:ssjs}
\end{figure}

\subsubsection{Active Server Pages (ASP)}
ASP is a server-side scripting environment for Web applications that
provides a scripting language engine that supports several languages,
an object model, and an interface for accessing server components. 
The most commonly used scripting language for ASP pages is VBScript. 
JScript, a JavaScript variant, is also supported out-of-the-box, and 
other languages can be installed. ASP pages are text files with {\tt .asp}
extension names that contain fixed HTML and blocks of scripting code 
within special brackets ({\tt $<${\%} .. {\%}$>$}) or {\tt $<$script$>$
.. $<$/script$>$} tag pairs.

The integrative power of ASP comes from the ability to access COM components 
on the server from within Web pages. The Component Object Model (COM) is a 
Microsoft standard that provides a language-independent means for defining 
and referencing components as objects. An ASP execution has built-in access 
to several automatically instantiated COM components, collectively known as 
the ASP built-in objects that have properties and methods that encapsulate 
the HTTP request-response cycle. ASP scripts can also reference the properties and 
methods of other COM components that are available on the Web server. A 
standard ASP installation includes a set of COM components that support 
common requirements of Web applications.  Figure~\ref{fig:asp} shows a simple 
ASP page that displays data from a table in a database.

\begin{figure}
\centering
\begin{tabbing}
{\em $<${\%} } \\
   \hspace*{2em}\= \+ {\em {\tt Dim} conn, rs} \\
   {\em {\tt Set} conn {\tt =} Server.CreateObject("ADODB.Connection")} \\
   {\em {\tt Set} rs {\tt =} Server.CreateObject("ADODB.Recordset")} \\
   {\em conn.Open "bookdb", "sa", "password" }\\
   {\em {\tt Set} rs {\tt =} conn.Execute("select au{\_}id, au{\_}fname, au{\_}lname from authors")} \-\\
{\em {\%}$>$} \\
$<$html$>$ \+\\
   $<$head$> \quad <$title$>$ASP Example Author Listing$<$/title$><$/head$>$ \\
   $<$body$>$ \\
      \hspace*{2em}\= \+ $<$h1$>$Author List$<$/h1$>$ \\
      $<$table$>$ \\
         \hspace*{2em}\= \+ $<$tr$><$th$>$ID$<$/th$><$th$>$First Name$<$/th$><$th$>$Last Name$<$/th$><$/tr$>$ \\
         {\em $<\% $ {\tt Do Until} rs.EOF {\%}$>$ } \\
            \hspace*{2em}\= \+ {\em $<$tr$><$td$><${\%}{\tt =}rs("au{\_}id") {\%}$><$/td$>$ } \\
            {\em $<$td$><${\%}{\tt =}rs("au{\_}fname") {\%}$><$/td$>$ } \\
            {\em $<$td$><${\%}{\tt =}rs("au{\_}lname") {\%}$><$/td$><$/tr$>$ } \\
            {\em $<\% $ rs.movenext } \\
            {\em {\tt Loop}} \-\\
         {\em {\%}$>$ } \-\\
      $</$table$>$ \-\\
   $</$body$>$ \-\\
$</$html$>$ 
\end{tabbing}
\caption{An example ASP page that retrieves data from a database table.}
\label{fig:asp}
\end{figure}

In the hands of experienced programmers, ASP's combination of HTML, 
scripting language code, and access to COM components can be a powerful tool 
for building dynamic Web applications. ASP became the most popular 
scripting-based templating technology on the Web, at least partially due to 
the large installed base of Microsoft server operating systems. The 
emergence of ASP was a step forward for Web application development; many 
systems supporting dynamic content that followed were influenced by the 
technology.

The disadvantages of ASP are related to portability, scalability, 
integration, reliability, and usability concerns.  Since ASP is still 
primarily applicable only to Microsoft systems despite various porting efforts, 
applications are generally not portable and integration with other platforms can 
be problematic. Scalability is limited due to runtime interpretation, which 
consumes extra CPU cycles compared to compiled code execution.  While COM 
integration provides access to an ad-hoc set of advanced features including 
support for transactions, the technology was not designed from ground up as a 
comprehensive, service-oriented framework, so critical concerns such as security 
and reliability are not integral to the environment. While the programming model 
is conceptually simple, the fundamental mismatch between the event-driven nature 
of applications and the page-centered interaction constraints of the Web is not 
addressed. The script-centered approach is less accessible to non-programmers 
than tag-centered approaches, so role separation are diminished.

\subsubsection{PHP}
PHP is an open-source, cross-platform, object-based scripting language 
for dynamic Web applications that is analogous to ASP. Versions of PHP run 
on many operating systems and Web servers, and interfaces are available to 
many database systems. PHP is available for Windows operating systems, but 
since ASP is generally the preferred option on Windows systems, PHP is 
most prevalent on Linux systems running Apache. The PHP syntax contains 
elements borrowed from C, Perl, and Java.  Figure~\ref{fig:php} shows an 
example PHP script.  While the advantages and disadvantages of PHP are
comparable to those of ASP, the portability of PHP can be beneficial if an 
application needs to run on several platforms or Web servers.  The open source 
combination of Linux, Apache, MySQL, and PHP, commonly referred to as LAMP, 
is gaining interest as a low cost platform for Web applications.

\begin{figure}
\centering
\begin{tabbing}
$<$html$>$ \\
   \hspace*{2em}\= \+ $<$head$>$ \\
      \hspace*{2em}\= \+ $<$title$>$PHP Example$<$/title$>$ \-\\
   $</$head$>$ \\
   $<$body$>$ \+\\
      {\em $<?$php } \\
         \hspace*{2em}\= \+ {\em {\$}res{\_}string {\tt =} ''; } \\
         {\em {\$}connval {\tt =} odbc{\_}connect ("bookdb", "sa",""); } \\
         {\em {\tt if} ({\$}connval) {\{} } \\
            \hspace*{2em}\= \+ {\em {\$}rs{\_}ret {\tt =} odbc{\_}exec({\$}connval,"select au{\_}lname + ', ' +  }\\
               \hspace*{2em}\= \+ {\em au{\_}fname as au{\_}name from authors"); } \-\\
            {\em {\tt if} ({\$}rs{\_}ret) {\{} } \+\\
               {\em {\tt echo} "The SQL statement executed successfully.$<$br$>$"; } \\
               {\em {\tt echo} "The results are below:$<$br$>$"; } \\             
               {\em {\tt echo} "$<$table$><$tr$><$td$><$b$>$Author Name$<$/b$><$/td$><$/tr$>$ "} \\
               {\em {\tt while} ({\$}res = odbc{\_}fetch{\_}row({\$}rs{\_}ret)) {\{} } \\
                  \hspace*{2em}\= \+ {\em {\$}res{\_}string {\tt =} } \\
                  \hspace*{2em}{\em "$<$tr$><$td$>$".odbc{\_}result({\$}rs{\_}ret,"au{\_}name")."$<$/td$><$/tr$>$" ; } \\
                  {\em {\tt echo} {\$}res{\_}string; } \-\\
               {\em {\}} } \-\\
           {\em {\}} } \\
           {\em {\tt else} {\{} } \+\\
              {\em {\tt echo} "The SQL statement did not execute successfully "; } \-\\
           {\em {\}} } \-\\
        {\em {\}} } \\
        {\em {\tt else} {\{} } \+\\
           {\em {\tt print}("$<$br$>$Connection Failed"); } \-\\
        {\em {\}} } \-\\
      {\em ?$>$ } \\
      $</$table$>$ \-\\
   $</$body$>$ \-\\
$</$html$>$ 
\end{tabbing}
\caption{An example PHP script that retrieves data from a database table.}
\label{fig:php}
\end{figure}

\section{SCALING UP TO THE ENTERPRISE}
\label{sec:enterprise}

In theory, server-side scripting environments such as ASP provide building 
blocks that support the needs of large enterprise systems, but in reality 
too much infrastructure responsibility is assigned to individual 
developers. Instead of being able to focus solely on business logic, Web 
application developers have to implement homegrown solutions for transaction 
management, resource pooling, and other complex features. For large-scale 
Web application development to be practical, infrastructure concerns need to 
be separated from business logic and presentation concerns.  

\subsection{Application Servers, Components, and Middleware}
\begin{figure}
\centerline{\includegraphics[height=6cm]{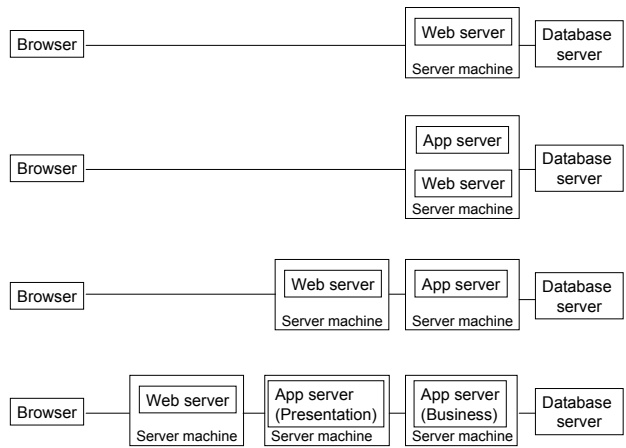}}
\caption{Example multi-tiered configurations for Web applications. In all of the
configurations an HTTP process and database server are required.} 
\label{fig:appServerTiers}
\end{figure}

Large-scale Web applications are typically built with an architecture 
that makes use of application servers and middleware to seperate the Web server,
presentation logic, business logic, and data access concerns into distinct tiers.
Figure~\ref{fig:appServerTiers} shows several common tier configurations 
(two-tier, three-tier, and four-tier) that are commonly used for enterprise 
Web applications.  The reliability and infrastructural requirements for large 
Web applications and the essential services that support the requirements, 
collectively known as {\em middleware services}, are summarized in 
Table~\ref{tab:middleware}.

\begin{table}
\begin{center}
\begin{tabular}{|p{0.6in}||l|p{2.4in}|} \hline
Requirement  & Approach/Solution & Description \\ \hline \hline
\multirow{6}{*}{Reliability}
   & Transparent fail-over & Route requests for failed services to another server. \\
   & Transaction support & Units of work completely succeed or are rolled back. \\
   & System management & Provide system monitoring and control capabilities. \\
   & High availability & Minimize time that a system is not available due to failures. \\
   & Replication & Duplicate resources for load balancing or recovery. \\
   & Failure recovery & Detect service failures, divert requests to another instance. \\ \hline
\multirow{7}{*}{Throughput} 
   & Load balancing & Alternate requests between servers to equalize utilization. \\ 
   & Clustering & Interconnect multiple servers to share a processing workload. \\
   & Threading & Execute requests concurrently within a process. \\
   & Efficiency & Process requests with minimal resources and latency. \\
   & Scalability & Maintain stable performance as the rate of requests increases. \\
   & Resource pooling & Share resources between multiple users in an optimized way. \\
   & Caching & Save computations for later compatible requests. \\ \hline
\multirow{7}{*}{Integration} 
   & Remote method invocation & Interfaces for synchronous method invocation. \\
   & Back-end integration & Interfaces to external information systems. \\
   & Database access & Store and retrieve database information. \\
   & Location transparency & Allow services to be requested by directory name. \\
   & Multi-protocol support & Integrate multiple protocols into a uniform interface. \\
   & Message-passing & Asynchronous communications though message-passing. \\
   & Legacy connection & Interfaces to obsolete or surpassed technologies. \\ \hline 
\multirow{2}{*}{Security} 
   & Logging and auditing & Record significant activities that occur within the system. \\
   & Permission checking & Verify identity and protect resources. \\ \hline
\multirow{8}{*}{Development} 
   & Rapid development & Reliable applications can be developed quickly. \\
   & Dynamic redeployment & Running applications can be updated without interruption. \\
   & Separation of concerns & Role-specific contributions are separate in code modules. \\ 
   & Modularity & Isolated changes minimally impact other parts of a system. \\
   & Reusability & Modules can be used for multiple applications. \\
   & Software scalability & Software remains manageable as system size increases. \\
   & Portable languages & Modules can run without changes on multiple platforms. \\
   & Standardization & The technology has multiple compliant providers. \\ \hline
\end{tabular}
\caption{Requirements for enterprise business systems.}
\label{tab:middleware}
\end{center}
\end{table}

The most widely applicable and effective middleware services are 
standardized, portable, and support rapid application development. 
{\em Application servers} aim to meet the middleware requirements 
of enterprise systems and are an 
integral part of large, dynamic Web applications. This section presents an 
overview of the standards, programming languages, and environments that 
support enterprise-class Web applications.

\subsection{Java-based Architectures}
Java is a popular, garbage-collected, object-oriented language developed by 
Sun Microsystems. Although Java initially gained attention on the Web as a 
client-side technology for running secure applets within browsers, its impact 
has ultimately been greater on the server-side. The portability of Java 
applications between platforms that support Java Virtual Machine (JVM) 
implementations is a particularly valuable feature for the Web, the most 
diverse computing environment imaginable. Reliability is improved relative to 
scripting languages because Java is statically type-safe. The environment 
supports runtime reflection, interface-implementation separation, and dynamic 
class loading, which are building blocks for components and application servers. 
None of the features are novel, but the packaging of the features within a 
single environment was timely and gained momentum with developers.

There have been two major versions, Java and Java 2. Java 2 de-emphasized applets 
and defined platform editions. A platform edition consists of a JVM, 
an SDK, and APIs. The Java Runtime Environment (JRE) consists of 
the JVM and components needed to run applications. Java 2, Standard Edition (J2SE) 
consists of the JRE and the SDK. The Java 2 platform, Enterprise Edition (J2EE) 
is the platform used for enterprise Web application development. J2EE includes 
J2SE and a set of specifications, designed to be implemented by application server 
providers, that defines a standards for enterprise middleware services that 
provide infrastructure for large scale Web applications.

\subsubsection{Java Servlets}
Java servlets extend Web servers to support dynamic content generation. The
Java Servlet API was introduced in 1997 as a replacement for CGI. Servlets 
are functionally similar to CGI scripts in that both are intermediary components 
that generate dynamic responses to HTTP requests. 

Problems areas for CGI that are addressed by servlets are performance, state 
management, and standardized access to middleware. The process model is 
similar to scalable CGI in that instances service multiple requests, but with 
servlets each instance is dedicated to a particular servlet. Servlets are executed 
by {\em servlet containers}, also known as {\em servlet engines}, which
are Java applications that manage threads through their lifecycle. The use of 
threads instead of processes improves efficiency and scalability, while providing 
efficient access to global objects initialized by the container. Since successive 
requests to a servlet are handled by a continuous thread, local objects 
instantiated within the thread are available throughout a servlet's
lifetime. Server-side state management is supported by the {\tt
ServletContext} and {\tt HttpSession} servlet API classes that hide
details of cookie writing, URL rewriting, and hidden field creation. Servlets 
have access to the large set of Java platform APIs.

The efficiency and scalability of servlets were the major factors that led to 
the emergence of Java as an important language for server-side Web programming, 
which is ongoing despite the insurgence of the .NET environment. The usability 
disadvantages of servlet programming are analogous to those of CGI. As with CGI, 
the servlet model provides only a thin abstraction over HTTP. Programmers must 
manually generate responses by writing to the standard output stream, so role 
separation is not well supported. Therefore, relatively few Web applications are 
composed entirely of servlets. Instead, servlets are used as an underlying mechanism 
behind template-based, model-driven, and framework-based technologies that 
were subsequently developed. 

\noindent
{\bf Servlet Containers.}
Servlet containers extend Web servers to implement the Java servlet API. 
A servlet container can function independently as a standalone Web server, or can 
be connected to an external Web server so that the container is dedicated to servlet 
processing. A request reaches a servlet container as the result of two mappings. The 
first mapping forwards requests from the Web server to the servlet container. The 
second mapping determines the servlet class to execute based on the servlet name
extracted from the request URI.  After a request is mapped, the actions taken depend 
on the current lifecycle stage of the servlet.  If the servlet is not active, the  
container loads the servlet class and creates a class instance, and invokes the servlet's 
{\tt init} method. Once the servlet instance is identified, the servlet container 
invokes the servlet's {\tt service} method to process the request, passing a {\tt request} 
object and a {\tt response} object. A {\tt request} object encapsulates information 
about HTTP requests, including parameters, attributes, headers, the request path, 
and the message body. A {\tt response} object encapsulates methods for building HTTP 
responses. Servlet developers override {\tt doGet} and {\tt doPost} methods that are 
invoked by the {\tt service} method to respond to HTTP request types 
({\tt GET} or {\tt POST}). The container calls a servlet's {\tt destroy} method when 
it needs to remove a servlet from service.

\noindent
{\bf Servlet Programming.}
Servlet programming is supported by classes and interfaces of the {\tt
javax.servlet} and {\tt javax.servlet.http} packages. A servlet must 
implement or extend a class that implements the {\tt Servlet}
interface. The {\tt Servlet} interface defines the lifecycle methods
of a servlet.  A typical servlet class extends {\tt HttpServlet} and
overrides the request processing methods as required by the
application. Figure~\ref{fig:servlet} shows an example of a very
simple Java servlet program.


\begin{figure}
\centering
\begin{tabbing}
{\tt import} com.test.search.*; \\
{\tt import} java.io.*; \\
{\tt import} javax.servlet.*; \\
{\tt import} javax.servlet.http.*; \\
\\
{\tt public class} ServletExample {\tt extends} HttpServlet {\{} \\
   \hspace*{2em}\= \+ String searchName {\tt =} null; \\ 
   SearchEngine searchEngine {\tt =} null; \\ 
   \\
   {\tt public void} init() {\tt throws} ServletException {\{} \\
      \hspace*{2em}\= \+ searchEngine {\tt = new} SearchEngine(); \-\\
   {\}} \\
   \\   
   {\tt public void} doGet(HttpServletRequest request, \+\\ 
      HttpServletResponse response) {\tt throws} IOException, ServletException {\{} \\
      response.setContentType("text/html"); \\
      PrintWriter out {\tt =} response.getWriter(); \\
      out.println("$<$html$><$head$><$title$>$Servlet Example$<$/title$><$/head$><$body$>$"); \\
      out.println("$<$h1$>$Enter a name to search for$<$/h1$>$"); \\ 
      \\
      {\tt if} ((searchName != null) {\{} \\
         \hspace*{2em}\= \+ out.println("The details are: "); \\
         String result {\tt =} searchEngine.search(searchName); \\
         out.println(result); \\
         searchName {\tt =} null; \-\\
      {\}} \\
      \\
      out.println("$<$p$><$form action=$\backslash $""); \\
      out.print(response.encodeURL("ServletExample")); \\
      out.print("$\backslash $" method=POST$>$"); \\
      out.println("$<$p$>$Search name: $<$input type=text size=30 name=$\backslash $"SearchName$\backslash $"$>$"); \\
      out.println("$<$p$><$input type=submit value=$\backslash $"Search$\backslash $"$>$"); \\
      out.println("$<$/form$><$/body$><$/html$>$"); \\
      out.close(); \-\\
   {\}} \\
   \\
   {\tt public void} doPost(HttpServletRequest request, \+\\
      HttpServletResponse response) {\tt throws} IOException, ServletException {\{} \\
      searchName {\tt =} request.getParameter("SearchName"); \\
      doGet(request, response); \-\\
   {\}} \-\\
{\}}
\end{tabbing}
\caption{An example Java servlet program.}
\label{fig:servlet}
\end{figure}

\subsubsection{JavaServer Pages}
JavaServer Pages (JSP) is a templating technology that was introduced
in 1999 to simplify servlet development.  An example JSP page is shown 
in Figure~\ref{fig:jsp}.  JSP pages are text files with {\tt .jsp} 
extension names that contain HTML, and Java code within special
delimiters (\textit{$<${\%} .. {\%}$>$}).  Java declarations can be
included within ({\tt $<${\%}! .. {\%}$>$}) delimiters as can
expressions within ({\tt $<${\%}= .. {\%}$>$}) delimiters. JSP translation 
directives can be included in JSP pages within ({\tt $<${\%}@ .. {\%}$>$}) 
delimiters.  Commonly used directives are {\tt @import}, used to import 
Java packages; {\tt @include}, used to include JSP fragment files into 
pages; and {\tt @taglib}, used to declare that a page uses a tag library.

The syntax diverges from the simplicity of ASP with the addition of elements 
that are closer in essence to SSI and ColdFusion. JSP pages can contain XML 
elements that initiate standard or custom execution phase {\em actions}. The 
standard actions defined in the JSP specification are recognizable as elements 
with {\tt jsp:} prefixes. Examples of standard actions are {\tt $<$jsp:usebean$>$},
{\tt $<$jsp:setproperty$>$}, and {\tt $<$jsp:getproperty$>$}, for
making use of JavaBean components; {\tt $<$jsp:include$>$}, used to
include static or dynamic resources within a page; and {\tt
$<$jsp:forward$>$} which transfers control to another Web page. Custom
actions invoke functionality in {\em tag libraries}, which are Java
components that are reused by JSP pages. Custom actions are
implemented by {\em tag files} with JSP syntax or by Java classes that
conform to with the JSP Tag Extension API.  The {\em JSP Standard Tag
Library (JSTL)} provides many tags supporting common functionality for
Web applications including conditional processing, iteration,
internationalization, XML manipulation, and database access. JSP 2.0
introduced the {\em Expression Language (EL)}, a simple language used
to embed expressions within JSP pages without Java scripting.

\noindent
{\bf JSP Containers.}
On the surface, JSP is reminiscent of ASP, but the implementation is 
very different. While ASP pages are interpreted, JSP pages are transformed 
by a JSP container into servlets.  A {\em JSP container} is a
modified servlet container that also serves JSP pages. JSP pages have
a lifecycle that includes translation (per page) and execution
(per request) phases. Translation can occur prior to deployment,
on deployment, or at runtime at the discretion of the implementer. In
the execution phase, the container routes requests to the servlet
thread that was created on behalf of a JSP page. The underlying
servlet mechanism is unchanged by the addition of the JSP technology,
so JSP can be considered to be an easier, alternative path to
developing servlets.

JSP is popular among developers because it is simpler than servlets 
for Web application development. At first glance, JSP appears to competently 
support role separation between designers and programmers. Simple examples 
like the JSP page shown in Figure~\ref{fig:jsp} appear to provide evidence. 
In practice, JSP pages tend to be dominated by Java code and XML tags which 
defeats role separation. Presentation is not adequately separated from 
behavior, leading to maintenance problems for large applications. The use of 
custom actions results in less code being directly placed in JSP pages and 
facilitates reuse, but does not address the essential problem; rather the 
problem is obscured since Java code is hidden behind action tags. Experience 
has shown that mixing Java and HTML produces systems that are hard to 
maintain ~\cite{Hunter:web:2000}. Since JSP was introduced, much of the 
subsequent innovation for Web development has been motivated by a search for 
techniques that properly exploit the capabilities of JSP and servlets.

\begin{figure}
\centering
\begin{tabbing}
{\em $<\% $@ {\tt page language}="java" {\tt import}="java.sql.*,ScheduleBean" {\%}$>$ } \\
{\em $<${\tt jsp:useBean} id="schedule" scope="session" class="ScheduleBean" /$>$ } \\
$<$html$>$ \\
   \hspace*{2em}\= \+ $<$head$>$ \\
      \hspace*{2em}\= \+ $<$title$>$Team Schedule$<$/title$>$ \-\\
   $</$head$>$ \\
   $<$body$>$ \+\\
      $<$p$>$Team Schedule:$<$/p$>$ \\
      $<$table$>$ \\
         \hspace*{2em}\= \+ $<$tr$>$ \\
            \hspace*{2em}\= \+ $<$td$><$b$>$Date:$<$/b$><$/td$>$ \\
            $<$td$><$b$>$Time:$<$/b$><$/td$>$ \\
            $<$td$><$b$>$Location:$<$/b$><$/td$>$ \\
            $<$td$><$b$>$Opponent:$<$/b$><$/td$>$ \-\\
         $</$tr$>$ \\
         {\em $<!--$ Get the appointment information --$>$ } \\
         {\em $<\% !$ ResultSet rs; {\%}$>$ } \\
         {\em $<\% $ schedule.setTeamName(request.getParameter("team")); } \+\\
            {\em rs {\tt =} schedule.executeQuery(); {\%}$>$ } \-\\
         {\em $<$\% } \+\\
            {\em {\tt while} (rs.next()){\{} } \-\\
         {\em {\%}$>$  }\\
            \hspace*{2em}\= \+ \hspace*{2em}\= \+ $<$tr$>$ \\
               \hspace*{2em}\= \+ {\em $<$td$><$b$><${\%}{\tt =} rs.getString("Date") {\%}$><$/b$><$/td$>$ } \\
               {\em $<$td$><$b$><${\%}{\tt =} rs.getString("Time") {\%}$><$/b$><$/td$>$ } \\
               {\em $<$td$><$b$><${\%}{\tt =} rs.getString("Location") {\%}$><$/b$><$/td$>$ } \\
               {\em $<$td$><$b$><${\%}{\tt =} rs.getString("Opponent") {\%}$><$/b$><$/td$>$ } \-\\
           {\em $</$tr$>$ } \-\-\\
        {\em {\%}$>$ } \+\\        
        {\em {\}} } \\
           {\em rs.close(); } \-\\
        {\em {\%}$>$ } \-\\
      $</$table$>$ \-\\
   $</$body$>$ \-\\
$</$html$>$
\end{tabbing}
\caption{An example of a simple JSP page.}
\label{fig:jsp}
\end{figure}

\subsubsection{Alternative Templating Engines}
Dedicated templating engines such as Velocity and WebMacro aim to replace 
JSP as a presentation technology for Web applications. Templating engines 
are generally designed to not allow Java code to be added to templates, so 
that developers are not tempted to mix business and presentation logic in 
templates. XML transformation, a notably different approach within the same 
problem space, is taken by Enhydra XMLC, currently an open source ObjectWeb 
project. With XMLC, plain XML templates are compiled into DOM template 
objects, which are Java classes that programmatically render the various 
kinds of output required by classes of clients, including HTML, WML, and 
other formats. Rather than embed Java code into the templates, programmers 
call XMLC-generated methods to manipulate properties of the DOM 
representation before the response is generated for the client. Apache 
Cocoon, another well-known XML transformation framework, supports dynamic 
content generation to multiple client types (channels) through its 
XSLT-based XSP templating language. 

\subsubsection{J2EE} 
\begin{figure}
\centerline{\includegraphics[height=6cm]{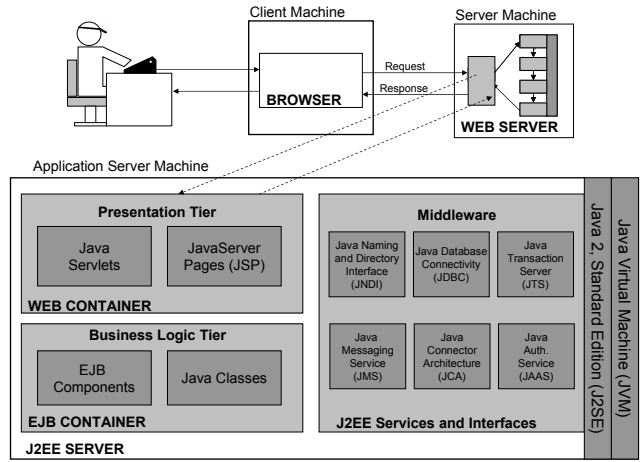}}
\caption{The J2EE architecture for Web applications.}
\label{fig:j2eeArch}
\end{figure}

The {\em Java 2 platform, Enterprise Edition (J2EE)} ~\cite{Roman:book:2001}
is a comprehensive set of 
standards designed to provide a portable environment that supports the 
requirements of large, enterprise Web applications. J2EE was first announced 
in April, 1997 and the first related specification, Enterprise JavaBeans 
1.0, was distributed in March, 1998. J2EE-compliant application servers were 
widely available by the end of 1999. Previous comparable technologies 
include transaction processing monitors, message-oriented middleware, and 
distributed object brokers (CORBA and COM). The older technologies generally 
only provided middleware services through explicit API calls. A component 
requiring transactional support had to make API calls in an appropriate 
sequence to begin and commit transactions. In contrast, J2EE supports 
{\em declarative middleware} by which containers provide middleware 
services (such as concurrency, 
transactional support, persistence, distribution, naming, and security) to 
components as specified in a set of declarative configurations. The use of 
declarative middleware allows developers to concentrate on business logic 
rather than on complex infrastructural details. The architecture of J2EE 
that supports Web applications is shown
in Figure~\ref{fig:j2eeArch}.

The most well-known J2EE specification is Enterprise JavaBeans (EJB), a 
complex standard for server-side components. An {\em EJB container} is 
an application server 
that provides an execution environment for EJB components. A J2EE server is 
an application server that implements all of the J2EE specifications. 
Table~\ref{tab:j2ee} summarizes the major J2EE specifications that are relevant for Web 
applications. When implementing a Java server-side Web application, it is 
important to decide which services are required for the application to help 
determine the class of application server that is needed. High-end 
application servers that implement all of the J2EE specifications tend to be 
much more expensive than simpler servers that provide fewer services. In 
many cases, support for servlets and JSP is sufficient to meet the needs of 
an application.

\begin{table}
\begin{center}
\begin{tabular}{|p{2.0in}||p{2.8in}|} \hline
Specification  & Description \\ \hline \hline
Enterprise JavaBeans (EJB) & Server-side managed Java components. \\ \hline 
Java Naming and Directory Interface (JNDI) & A directory interface for locating components. \\ \hline
Java Database Connectivity (JDBC) & An interface for accessing relational databases. \\ \hline
Java Transaction Server (JTS) & Declarative transactional support for EJB components. \\ \hline
Java Messaging Service (JMS) & Allows components to communicate through messaging. \\ \hline
Remote Method Invocation (RMI) & An interface for communication between distributed objects. \\ \hline
Java Servlets and JavaServer Pages (JSP) & Web server extension components. \\ \hline 
JavaMail & Allows email to be sent from Java programs. \\ \hline
J2EE Connector Architecture (JCA) & Connectivity to legacy systems. \\ \hline
Java API for XML Parsing (JAXP) & XML parsing \\ \hline
Java Authentication and Authorization Service (JAAS) & Security for EJB components. \\ \hline
\end{tabular}
\caption{J2EE specifications.}
\label{tab:j2ee}
\end{center}
\end{table}

J2EE has been successful and compliant application servers are
available from many providers. The platform can be very scalable if
properly exploited, but the size and complexity of the architecture
can make it a difficult environment to work with. The learning curve
for J2EE developers can be relatively lengthy, especially if EJB is
used, due to the complexity of the component architecture.

\subsection{.NET}
.NET is an enterprise computing platform which is provided as a set of
products for Microsoft operating systems.  The features of .NET are 
comparable to those of J2EE. While designed to be a cross-platform 
environment, .NET is most relevant for Microsoft operating systems, although 
the open source Mono project promises to bring .NET to Linux. Due to the 
large installed base of Microsoft operating systems, .NET has been steadily 
gaining traction with developers since its initial production release in 2001.

\begin{figure}
\centerline{\includegraphics[height=6cm]{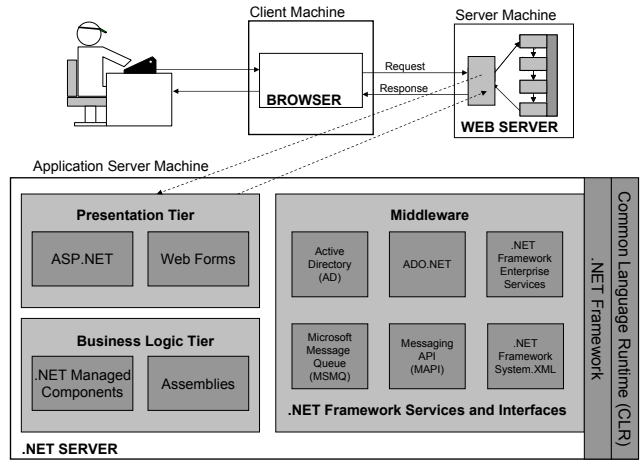}}
\caption{The .NET architecture for Web applications.}
\label{fig:dotnetArch}
\end{figure}

The main components of the .NET platform are the {\em Common Language Runtime (CLR)} and 
the {\em .NET Framework} class library.  The CLR is a virtual machine that dynamically 
compiles {\em Microsoft Intermediate Language (MSIL)} byte code into native executable 
code at runtime. .NET is a multi-language environment in that multiple source languages 
can be compiled into MSIL byte code and executed by the CLR. The {\em Common Type System 
(CTS)} defines how types are declared and used in the CLR, providing a basis for type 
interoperability between modules implemented in different languages. Self-describing 
components, known as {\tt assemblies} within .NET, are managed and executed by the CLR. 
The .NET Framework class library is a large class library that provides similar functionality 
as the Java Platform APIs. Assemblies and the types they define are hierarchically grouped 
into namespaces that can be referenced by programs. While many programming languages are 
supported, the primary development language is C$^{\# }$, which is very similar to Java.

.NET provides an environment for enterprise Web applications that is 
comparable to J2EE. Table~\ref{tab:j2eenet} compares the features of the two 
environments. ASP.NET is a reworked version of ASP that enables 
rapid development of Web applications that make use of the capabilities of 
the .NET framework.  The architecture of .NET for Web applications is shown
in Figure~\ref{fig:dotnetArch}.

\begin{table}
\begin{center}
\begin{tabular}{|p{2.0in}|p{2.7in}|} \hline
J2EE Feature & .NET Equivalent \\ \hline \hline
Java & Primarily C$^{\# }$, other languages are supported \\ \hline 
Java Virtual Machine (JVM) & Common Language Runtime (CLR) \\ \hline 
Enterprise JavaBeans (EJB) & Assemblies, .NET CLR managed components \\ \hline 
Java Naming and Directory Interface (JNDI) & Active Directory (AD) \\ \hline 
Java Database Connectivity (JDBC) & ADO.NET \\ \hline 
Java Transaction Server (JTS) & COM+, .NET Framework System.EnterpriseServices \\ \hline 
Java Messaging Service (JMS) & Microsoft Message Queue (MSMQ) \\ \hline 
Remote Method Invocation (RMI) & Remote Method Invocation (RMI) \\ \hline 
Swing API & Win Forms \\ \hline 
Java Servlets and JavaServer Pages (JSP) & ASP.NET, Web Forms \\ \hline 
Java Server Faces (JSF) & ASP.NET, Web Forms \\ \hline 
JavaMail & Messaging API (MAPI) \\ \hline 
Java API for XML Parsing (JAXP) & .NET Framework System.XML \\ \hline 
\end{tabular}
\caption{J2EE - .NET platform feature cross-reference.}
\label{tab:j2eenet}
\end{center}
\end{table}

\subsection{Architectures Based on Other Languages}
While J2EE and .NET are the most well-known platforms for enterprise
Web applications, alternatives exist.  Python is a simple, portable, 
freely available object-oriented programming language that has been
used as a basis for enterprise systems development ~\cite{Trauring:eaij:2003}.  
Twisted is a large, stable framework that provides support for 
distributed and 
networked applications as well as Web applications ~\cite{Twisted:web:2005}.
The breadth of Twisted illustrates the difficulties involved in supporting
enterprise Web development outside of J2EE and .NET: the framework
includes a Web server and a DNS server; enterprise capabilities including
authentication and database connectivity; a distributed-object broker;
protocol abstractions and implementations including HTTP, SMTP, IRC, 
DNS, Telnet, TCP, and UDP; and interoperability with 
Python toolkits ~\cite{Vinoski:ieeeic:2004}.  Another well-known Python
framework that is similar to Twisted is the Python Enterprise 
Application Kit (PEAK) ~\cite{PEAK:web:2005}. Other languages that
have been proposed for enterprise web development are 
Curl ~\cite{Ward:ijwet:2003}, Croquet ~\cite{conf:c5:Smith:2003},
Haskell ~\cite{Meijer:jfp:2000}, LISP ~\cite{Graham:web:2001}, 
Perl ~\cite{Rolsky:book:2003}, Scheme ~\cite{conf:afp:Felleisen:2002}, 
and Smalltalk ~\cite{Bryant:web:2004}.  Each
alternative language and environment faces an uphill road on the way towards
widespread adoption in enterprises due to the popularity of J2EE
and .NET, while also facing the difficulties of either reimplementing
basic Web functionality or relying on prexisiting, low-performing 
features of the infrastructure such as CGI.

\section{CLASSIFICATION OF SYSTEMS SUPPORTING DYNAMIC WEB CONTENT}
\label{sec:classification}

This section summarizes the survey so far by presenting a
classification of technologies that are relevant to dynamic content
generation for the Web. Reflective of the major system requirements
for dynamic Web applications, the top level of the classification is
divided into foundational, integration, and dynamic user interface
generation technology classes.  Each top level class is sub-classified
by technology type, form, and function to yield comparable groups.  At
the most basic level of the classification, key properties of
well-known examplars are named and compared with the objective of
providing technology selection guidelines for Web development
projects.

{\bf Foundational Technologies.}  The basic functionality of Web
browser clients is defined by the HTTP protocol and content
description language standards including HTML, XHTML, and CSS.  Web
server functionality is also defined by the HTTP protocol.
Table~\ref{tab:foundational} classifies the foundational technologies
of the Web. Table~\ref{tab:browserserver} shows the most commonly used
Web browsers and Web servers based on market share as of January,
2005.

\begin{table}
\begin{center}
\begin{tabular}{|l|l|p{2.0in}|} \hline
Technology & Tier & Implementations \\ \hline \hline
Markup Standards & Client & HTML, XHTML, CSS \\ \hline
Protocol specifications & Server & HTTP, SSL, MIME, WebDAV \\ \hline
Web browser & Client & Internet Explorer, Opera, Mozilla, Netscape \\ \hline
Web server & Server & Apache, Internet Information Server, 
Sun Java Web Server, Zeus Web Server, Jigsaw \\ \hline
\end{tabular}
\caption{Foundational technologies for the Web.}
\label{tab:foundational}
\end{center}
\end{table}

\begin{table}
\begin{center}
\begin{tabular}{|p{0.7in}||p{2.0in}|p{1.3in}|} \hline
Tier & Product (Provider) & Platform (Tier share {\%}) \\ \hline \hline
\multirow{4}{*}{Web browsers}  
   & Internet Explorer (Microsoft) & Windows (69.7) \\ 
   & Opera (Opera Software) & Several (1.9) \\ 
   & Mozilla (The Mozilla Organization) & Several (23.3) \\ 
   & Netscape (AOL / Netscape) & Several (1.4) \\ \hline
\multirow{4}{*}{Web servers}  
   & Apache (Apache Software Foundation) & Several (68.4) \\ 
   & Internet Information Server (Microsoft)	& Windows (20.9) \\ 
   & Sun Java System Web Server (Sun) & Several (23.3) \\ 
   & Zeus Web Server (Zeus Technology) & Several (1.2) \\ \hline
\end{tabular}
\caption{Web browsers and servers.}
\label{tab:browserserver}
\end{center}
\end{table}

{\bf Integration Technologies.}
Integration technologies do not directly generate dynamic content.   
Rather they provide interfaces that allow web sites to access information 
in other systems.  Table~\ref{tab:clientintegration} provides a 
classification of client-based integration technologies.   
Table~\ref{tab:serverintegration} shows the most commonly-used
server-side integration technologies.

\begin{table}
\begin{center}
\begin{tabular}{|p{1.1in}||p{1.4in}|p{2.2in}|} \hline
Technology & Exemplars & Key Properties and Common Usage \\ \hline \hline
{\bf Browser interfaces} & \multicolumn{2}{|l|}{} \\ \hline
State management & Netscape Cookie API 
   & Used for client state persistence. \\ \hline 
Web service client interfaces & SOAP, XML-RPC 
   & Highly portable. Increased interactivity. \\ \hline
Scripting interfaces & XMLHttpClient 
   & Highly portable.  Increased interactivity. \\ \hline
\end{tabular}
\caption{Client-based integration technologies.}
\label{tab:clientintegration}
\end{center}
\end{table}

\begin{table}
\begin{center}
\begin{tabular}{|p{1.1in}||p{1.4in}|p{2.2in}|} \hline
Technology & Exemplars & Key Properties and Common Usage \\ \hline \hline
{\bf Programming languages} & \multicolumn{2}{l|}{} \\ \hline
   Natively compiled	& C, C++ 
      & Highly portable, high performance, low usability. CGI scripting and web server extension. \\ \hline
   Interpreted &	Perl, Python, Ruby  
      & Highly portable.  Performance varies.  CGI scripting. \\ \hline
   Byte-code compiled & Java, .NET languages 
      & Highly portable.  Good performance.  Extensive language run-time libraries. \\ \hline
{\bf Components} & \multicolumn{2}{|l|}{} \\ \hline
   Complex components & CORBA, COM, EJB, .NET managed components
      & Enterprise systems development.  \\ \hline
   Simple components & JavaBeans, Spring, picoContainer, Java classes
      & Enterprise systems development.  Improved usability.  \\ \hline
{\bf Middleware} & \multicolumn{2}{|l|}{} \\ \hline
   XML transformation & XSLT, Cocoon	
      & XML Web publishing.  \\ \hline
   Messaging	& MSMQ, MQSeries, J2EE JMS 
      & Enterprise systems development. \\ \hline
   Security & J2EE JAAS 
      & Enterprise systems development.  \\ \hline
   Data access & ODBC, JDBC, ADO.NET 
      & Highly portable.  Highly scalable.  \\ \hline 
   Object-relational mapping &	Hibernate, JDO, TopView, iBatis  
      & Highly portable.  Highly scalable.  Improved usability.  \\ \hline
   Web services &	SOAP, XML-RPC  
      & Highly portable. \\ \hline
\end{tabular}
\caption{Server-side integration technologies.}
\label{tab:serverintegration}
\end{center}
\end{table}

{\bf Dynamic Content Generation Technologies.}
Table~\ref{tab:clientdynamic} provides a classification of dynamic 
technologies for Web clients. Table~\ref{tab:serverdynamic} provides 
a classification of dynamic technologies for Web servers.  

\begin{table}
\begin{center}
\begin{tabular}{|p{1.1in}||p{1.4in}|p{2.2in}|} \hline
Technology & Exemplars & Key Properties and Common Usage \\ \hline \hline
{\bf Browser extension} & \multicolumn{2}{l|}{} \\ \hline
   Browser specific APIs & CCI, ActiveX, Netscape Plug-API  
      & High performance, low usability.  Extend browser capabilities. \\ \hline
{\bf Client dynamism} & \multicolumn{2}{|l|}{} \\ \hline
   Directly interpreted & JavaScript, VBScript, DOM, SVG
      & Highly portable due to Web standards acceptance. Increased interactivity. \\ \hline
   Byte-code & Macromedia Flash, Java applets 
      & Highly portable depending on plug-in spread. Marketing presentation development. \\ \hline
{\bf Rich interfaces} & \multicolumn{2}{|l|}{} \\ \hline
   Browser alternatives & CURL, Sash Weblications, Konfabulator, XAML 
      & Used for applications with high interactivity 
        requirements that need to access the Internet. \\ \hline
   Client frameworks & Jakarta Commons httpClient 
      & Used for browser alternative development. \\ \hline
   Forms interfaces & InfoPath, XForms 
      & Used by forms-based business systems. \\ \hline
   Dynamic assembly & Edge Side Includes (ESI), Client Side Includes (CSI) 
      & Improved cached content delivery. \\ \hline
\end{tabular}
\caption{Client-based dynamic content generation technologies.}
\label{tab:clientdynamic}
\end{center}
\end{table}

\begin{table}
\begin{center}
\begin{tabular}{|p{1.1in}||p{1.4in}|p{2.2in}|} \hline
Technology & Exemplars & Key Properties and Common Usage \\ \hline \hline
{\bf Server extension} & \multicolumn{2}{l|}{} \\ \hline
   Server-specific & ISAPI, NSAPI, Apache API 
      & Proprietary. Scalable.  Complex.  Implement extended server capabilities. \\ \hline
{\bf Gateways} & \multicolumn{2}{|l|}{} \\ \hline
   Simple & CGI 
      & Portable.  Not scalable. Low usability.  
        Small-scale applications with simple navigational requirements. \\ \hline
   Scalable & Fast CGI 
      & Portable.  Scalable. Low usability.  
        Medium to large-scale applications with simple navigation requirements. \\ \hline
{\bf Interpreters} & \multicolumn{2}{|l|}{} \\ \hline
   General purpose & Server-side JavaScript, mod-perl 
      & Medium to large-scale applications with simple navigation requirements. \\ \hline
   Template-based & SSI, XSSI, ColdFusion, ASP, PHP, JWIG 
      & Portable.  Medium to large-scale applications with simple navigation requirements. \\ \hline
{\bf Extended servers} & \multicolumn{2}{|l|}{} \\ \hline
   Servlet engines & Tomcat, Resin 
      & Highly portable.  Scalable.  Medium to enterprise-scale 
      applications. Frameworks.  \\ \hline
   Template engines & JSP, Velocity, WebMacro, XMLC, FreeMarker 
      & Portable.  Scalable.  Generally used as the view 
        component of MVC implementations. \\ \hline
   Content management & Zope, Cocoon 
      & Scalable.  Web publishing. \\ \hline
   J2EE application servers & WebSphere, jRun, JBOSS, WebLogic 
      & Highly portable.  Highly scalable.  Complex. Enterprise systems. \\ \hline
\end{tabular}
\caption{Server-side dynamic content generation technologies.}
\label{tab:serverdynamic}
\end{center}
\end{table}

\section{WEB PROGRAMMING VS. REGULAR PROGRAMMING}
\label{sec:programming}

In software development terms, the maturity level of the state of common 
practices for Web development has traditionally lagged relative to the 
technologies and techniques used for other client-server applications
~\cite{Gellersen:ieeeic:1999}.  As late as 1995, the CGI was still the 
most practical option for dynamic Web content creation.  In contrast, 
distributed object environments based on CORBA and COM have been available 
for client-server development since 1992. By the time that templating and 
scripting languages were commonly supporting largely ad-hoc Web development 
practices, client-server development more advanced, supported by graphical 
development tools, frameworks, and software engineering practices.

As the Web began to be used in an increasingly large class of critical
business applications, it became apparent that the fundamental
requirements were not well supported by existing solutions.  Early
attempts, roughly between 1995 and 1999, centered on trying to find
a unifying API for Web programming, essentially viewing the Web as a
distributed object system in the tradition of CORBA
~\cite{W3C:web:1997,Cardelli:ieeetse:1999,Manola:ieeeic:1999,Thompson:acmcsur:1999}.
Difficulties in coordinating the efforts of the wide-ranging Web community
hindered efforts to define a global API, but the mark of distributed object 
research can be seen in {\em service-oriented architecture (SOA)} standards, 
which implement globally distributed Web service technologies by exchanging XML 
over HTTP ~\cite{W3C:web:2004:b}.

The continual disparity fueled a marketing pipeline for dynamic Web 
technology creators. Almost any advance that addressed limitations of 
Web development could find a waiting base of potential adopters.  Even 
problematic technologies such as ActiveX found avenues of acceptance 
solely based on incremental benefits.  The introduction of J2EE in 
1999 was a flashpoint for the dynamic Web; instantly the maturity gap was 
narrowed and priorities shifted so that many software engineering advances 
were for the first time being driven by the requirements of Web applications.  
The importance of J2EE can not be overstated since it set standards that 
have since influenced subsequent significant advances for Web development, 
including .NET, which surfaced as a competitive response. 

This section examines tools, techniques, and technologies that have been 
carried forward from traditional programming domains into the realm of the
Web.

\subsection{Frameworks and Patterns}
While J2EE and .NET provide infrastructure needed to build reliable 
enterprise-scale Web applications, the responsibility for effectively 
using the technology falls on individual implementers. Many Web 
applications are built in an ad-hoc fashion, without regard for software 
engineering principles, deeply entangling content, presentation, and 
behavior so that the structure of the implementation is obscured
~\cite{Pressman:ieeesw:2000}. Recently much effort has been devoted to 
devising methods that exploit the technology base in a disciplined way 
while simplifying development. The tangible result has been the emergence 
of a large number of Web application development frameworks. The Java 
community has been very influential in that the most popular frameworks 
are open source Java projects.  Well-known Web application frameworks 
are also available for server-side scripting languages such as Perl, PHP, 
and Python. While ASP.NET and the .NET framework provide a wealth of support
for Web development out-of-the-box, .NET is also becoming a target for 
framework development. This section discusses frameworks in relation to 
Web application development along the requirements and patterns that 
prompt their development.

\subsubsection{Characteristics of Frameworks} 
Frameworks are extensible module sets supporting rapid development of 
repetetive requirements in a reliable way though reusable skeletal design 
pattern realizations.  The objective is to allow developers to concentrate
on the problem domain rather than on low-level implementation details,
while providing concrete benefits beyond the associated acquistion and 
learning curve costs.  Frameworks are comparable in terms of
documentation quality, extensibility, modularity, evolvability, maturity, 
and white/black box properties ~\cite{Fayad:cacm:1997}. Poorly documented
and immature frameworks should be avoided for critical applications.

\subsubsection{Categories of Web Frameworks} 
The most useful frameworks supporting Web application development can be 
categorized as Web application and user interface frameworks, persistence 
frameworks, and lightweight containers ~\cite{Nash:book:2003}. Web application 
and user interface frameworks are the most directly relevant to dynamic content
generation systems. Persistence frameworks such as Hibernate aim to allow 
programmers to efficiently retrieve and update database information through 
encapsulated objects rather than through direct SQL calls, or through EJB 
entity beans.  Frameworks that leverage inversion of control to simplify 
component-based development, collectively known as {\em lightweight containers}, 
are gaining traction as an more usuable alternative for J2EE applications that 
would otherwise needlessly incur the implementation complexity of EJB, even 
though high-end capabilities are not required. The open source PicoContainer 
and Spring frameworks are highly-regarded lightweight container frameworks 
~\cite{Johnson:book:2004}.

\subsubsection{Model-View-Controller}
\begin{figure}
\centerline{\includegraphics[height=5cm]{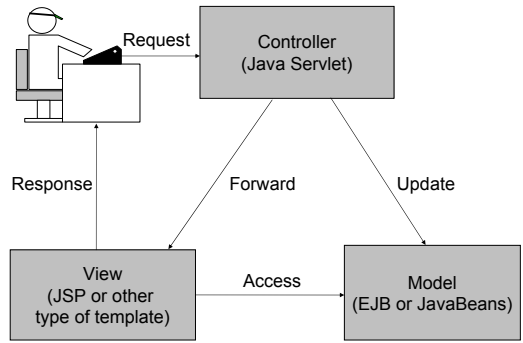}}
\caption{Adapting the MVC architecture for the J2EE Web applications.}
\label{fig:mvcMappingJ2EE}
\end{figure}

The {\em Model-View-Controller (MVC)} design pattern 
~\cite{Krasner:joop:1988}, commonly used in 
user interface programming to separate presentation, business, and state 
management concerns, is a logical architectural choice that matches the 
event-driven nature of dynamic Web applications.  Early Web technologies 
did not allow convenient separation of concerns, but the more recent 
convergence of Java, scripting and template languages, and servlets 
supports the modularization of Web applications to align with the MVC 
roles.  Figure~\ref{fig:mvcMappingJ2EE} shows the most common mapping of 
the MVC roles to J2EE entities.  For .NET, a similar mapping to ASP.NET, 
the IHttpHandler interface, and managed components is possible. Much 
attention has been focused on creating Web MVC frameworks since 2000,
when the potential for reuse and streamlining the development 
process became evident.  As a result, many competing frameworks are 
available, many of which are products of open source development projects.  
While the most well-known frameworks currently target the J2EE platform,
several have been ported to .NET, and there are also many scripting 
language frameworks available.  Scripting languages frameworks  
are handicapped from the start by either a reliance on CGI or the need to 
implement a supporting infrastructure analogous to servlets, in addition to
supporting MVC.  

\subsubsection{Application-driven Web MVC Frameworks}
Application-driven Web MVC frameworks implement MVC using 
the {\em Front Controller} pattern ~\cite{Fowler:book:2002}. In the Front 
Controller pattern, events are directed to an application-level Controller 
that invokes the correct action in response. The event-action table is 
is maintained at the application level, usually in an XML file.  Navigation 
details are abstracted out of individual pages, although the encapsulation
and reusability is compromised by dependencies on the configuation file 
that defines the event-action table. An application consists of a Controller 
class, Model classes, View page templates, and a configuration file. The main 
objective is event-handling rather than hiding implementation details, so 
programmer familiarity with resource implementation technologies is assumed. 
The lightweight nature of the abstraction reduces the learning curve for 
experienced Web developers relative to more opaque frameworks.  Novice Web 
programmers may face an extended orientation period due to the need to 
comprehend the workings of a framework in addition to more fundamental 
concepts.

\noindent
{\bf Apache Struts.}
The most well-known application-driven Web MVC framework is the open 
source {\em Apache Struts} framework ~\cite{Husted:book:2003}. First 
introduced in 2000, Struts continues to be the most popular Web MVC 
framework for Java.  The framework is mature, well-documented, and 
effectively supports the requirements of a large class of interactive 
Web applications. The implementation is straightforward and  
based on the servlet, JSP, HTML forms, JavaBeans, and XML standards. 
While several well-known frameworks with active developer
communities, including WebWork, Spring MVC, and Maverick, occupy the 
same architectural niche, the details of Struts, the de-facto
leader, are broadly representative of the category.

The event-action table for a Struts application is specified in 
{\tt struts-}{{\tt config.xml}, an application-level configuration 
file. The {\tt web.xml} configuration file for an application maps 
URI names to the {\tt ActionServlet} servlet, a central Controller 
servlet supplied by the framework that routes requests to {\tt Action} 
class instances that encapsulate response logic. {\tt Action} instances 
access session data, HTML form data, and integration components to 
formulate dynamic responses. Form beans, JavaBean components that 
implement the Struts {\tt ActionForm} or {\tt DynaActionForm} interfaces, 
encapsulate the server-side state of the input fields of an HTML form. 
The {\tt DynaActionForm} interface simplifies state management by 
dynamically creating form beans when they are needed without requiring
additional programming. After processing response logic, an 
{\tt Action} class transfers control to the Controller passing a
{\em global forward}, the logical name of the View template that will 
generate the next page of the dynamic user interface based on the outcome
of the response logic.  Although View templates are normally JSP pages, 
the framework allows other template engines to be used, including
Velocity and WebMacro.  The Model is the least constrained tier of 
a Struts application, consisting of components that are accessed by 
{\tt Action} class instances and View templates to dynamically access and
update persistent information. Field-level input validations, 
implemented either by hand-coding form bean {\tt validate()} methods 
or by declaration in the {\tt validator-}{\tt rules.xml} and 
{\tt validation.xml} configuration files, are always performed on 
the server-side and optionally through generated JavaScript on the 
client-side.

\subsubsection{Page-driven MVC Frameworks}
Page-driven frameworks implement the {\em Page Controller} 
pattern ~\cite{Fowler:book:2002} to provide an event-driven model for 
Web programming that recalls traditional desktop GUI programming. In the Page 
Controller pattern, events generated from pages are directed to page-level 
Controllers. The event-action table is spread throughout individual 
pages of an application. An application consists of related pages, classes, 
and configuration files.  Relative to application-driven frameworks, the higher 
degree of page independence allows heavier abstraction over resource 
implementation details that supports rapid component-based development 
through drag-and-drop GUI page composition.  The high abstraction level
may extend the learning even for experienced Web programmers due to the need 
to become familiar with with a completely different object model. 

\noindent
{\bf Desktop API Derivatives.}
Several Java page-driven frameworks, including the open source Echo 
~\cite{Nextapp:manual:2005} and wingS ~\cite{wings:manual:2005} frameworks, 
implement the object model of the Java Swing API to literally apply desktop 
GUI programming techniques to Web development. WebCream 
~\cite{Creamtec:manual:2005} takes the approach to its logical conclusion 
by converting compiled Swing applications into HTML pages at the 
Web server dynamically at runtime.  The effectiveness of desktop API-derivative 
Web application frameworks is limited by their code intensive nature, 
which prevents role separation between designers and programmers. GUI development 
tools such as EchoStudio simplify development, but are not accessible to Web 
designers since page designs are not based on HTML so familiar Web authoring 
tools can not be used.

\noindent
{\bf WebObjects.}
Other page-driven frameworks take a more practical, template-based approach 
to incorporating desktop GUI programming practices into Web development. The 
proprietary Web development framework of the NeXT (now Apple) WebObjects 
application server environment pioneered a page-driven, component-based 
approach to Web programming in 1996. The WebObjects framework was initially 
built for Objective-C, but was re-implemented for Java in 2000 to support 
J2EE development. WebObjects applications are collections of pages containing 
HTML and references to {\em Web Components}.

\noindent
{\bf Tapestry.}
Tapestry ~\cite{Ship:book:2004}, available since 2000, is an open source Java 
framework from Apache that was heavily influenced by WebObjects.  Tapestry 
provides an object model for component-based development of Web applications. 
A Tapestry application is a collection of pages that are composed from HTML and 
components that encapsulate dynamic behavior. In Tapestry, components are known 
as {\em Java Web Components (JWC)}. Tapestry supports two kinds of components, 
user interface components and control components. Control components are not 
rendered on pages but instead provide control flow constructs. Simple 
applications can be constructed entirely from library components provided as 
part of the framework distribution. A Tapestry page is defined by an 
XML specification, one or more Java classes, and an HTML template. The XML 
specification identifies a Java page controller class, and defines identifiers 
that indirectly bind components to HTML templates. The page controller class 
implements {\tt listener} methods that handle user interface events. Templates 
contain HTML and component references. A Tapestry component definition includes 
an XML specification, one or more Java classes, and an HTML template. Both page 
and component templates consist of plain HTML and placeholder tags that reference 
components through a special attribute, {\tt jwcid}. Role separation is 
well-supported since Web designers can use their preferred authoring tools to 
design page templates, which templates contain only valid HTML. Although the 
framework internally routes all requests through a single entry servlet, the 
{\tt ApplicationServlet}, the Tapestry object model completely abstracts servlet 
processing.  Programmers do not need to understand servlet processing to 
effectively use the framework. Tapestry appeals to desktop GUI programmers 
since they are familiar with event-driven programming. While the dynamic content
generation process is computationally intensive, the framework avoids scalability 
problems by efficiently caching internal objects. 

\noindent
{\bf ASP.NET and JavaServer Faces.}
While Tapestry is technically highly-regarded, the momentum behind the 
framework has been largely eclipsed by the emergence of ASP.NET 
~\cite{Esposito:web:2003} and, to a lesser extent so far, the nascent JavaServer 
Faces (JSF) specification ~\cite{Mann:book:2004}. ASP.NET is 
an upgraded version of ASP that supports Web Forms, a namespace 
within the .NET framework that provides a page-driven object model for Web 
programming that is similar to the Tapestry object model. JSF is a 
specification for a component-based Web application framework built over JSP 
and tag libraries, much closer in concept to Struts than ASP.NET. JSF 
has superficial similarities to ASP.NET, but is very different in detail. 
Both frameworks support rapid user interface development with GUI form 
builder tools, primarily Visual Studio.NET for ASP.NET and Java Studio 
Creator for JSF. A major conceptual difference is that ASP.NET is 
page-driven, while JSF is application-driven. All requests for JSF 
application resources are routed to views by a central servlet, the 
{\tt FacesServlet}, per specifications in the application-level 
{\tt faces-config.xml} file. While the uptake of JSF is in an early stage 
and widespread adoption is not inevitable, the merging of characteristics 
of the Front Controller and Page Controller patterns provides a higher degree 
of deployment flexibility due to clearer separation of the navigational aspect 
from page definitions relative to ASP.NET.

\noindent
{\bf Portals and Portlets.}
The component model of {\em portlets} is closely related to JSF, which
features integration with the Java Portlet API ~\cite{Abdelnur:web:2003}. 
Portlets are managed Java components that respond to requests and generate 
dynamic content. Java Portlet API specifies how to compose component portlets 
into combined {\em portals} that aggregate content from portlet subject to
personalization.  A similar component model is provided by the ASP.NET 
WebParts framework.

\subsection{Other Programming Technologies}

Several technologies have been developed that attempt to address
the mismatch between flow control in Web development relative to 
other kinds of programming. The disconnect is mainly due to the fact 
that user interface processing is distributed in Web programs, 
with clients displaying interfaces that are at least partially formulated 
on servers.   Functional programming with continuations has been used 
to generate dynamic content with some success, but remains outside of 
the mainstream.  Continuations are used to abstract the retention of 
state information between interactions.
The MAWL, $<$bigwig$>$, JWIG, and Cocoon Flow projects implement 
{\em session-centered programming}, a method for combining
page template fragments with control flow logic that maintains the
state of local variable by a mechanism that has similarities to
continuations. 

\subsection{Model-Driven Development of Web Applications}

The defining feature of model-driven development is automatic code
generation of deployable applications from high-level feature
specifications. Model-driven development technologies for Web
applications aim to simplify the development process by generating
deployable sites from presentational, behavioral, and navigational
requirements specified in models. In the tradition of prior work
in automatic programming and CASE, which were not completely
successful, model-driven development technologies aim to reduce the
dependency on low-level programming by raising the abstraction model
to a higher level. Adaptation to Web development required the creation
of new kinds of models, methods, and techniques that better match the
unique properties of Web applications.

\noindent
{\bf Initial Progress.}
Araneus ~\cite{Atzeni:vldb:1997} and Strudel 
~\cite{Fernandez:dsl:1999}, are representative of initial
research progress in adapting model-driven techniques for Web development.
These systems utilize data models to manage collections of generated content derived 
using database metadata and queries. ~\citeN{Fraternali99:acmcsur:1999}
surveyed current work in the area as of 1999, including AutoWeb,
RMM, OOHDM, and the Oracle Web Development Suite, each of which
applied proprietary database, navigation, behavior, and presentation
modeling approaches to Web development. WebML and its commercial
successor, WebRatio, use proprietary hypertext, data, and presentation 
models to comprehensively extend the prior work by generating code for an 
abstract framework that maps to platform-specific MVC implementations at 
runtime.  Other products, including CodeCharge, CodeSmith, DeKlarit, and 
Fabrique, emphasize GUI-based maintenance of detailed models that 
facilitate generative programming of the presentation tier for Web applications, 
at varying degrees of rigor.  While these initial approaches were workable, 
widespread usage of the technologies was limited by the reliance on proprietary 
modeling languages.
 
\noindent
{\bf Model-driven Architecture.} The Model Driven Architecture (MDA)
standard from OMG chose UML as the primary modeling
language for model-driven development. UML was formally extended to be
computationally complete in order to be able to support the level of
detail needed to generate applications of abitrary complexity from
specifications. The MDA standards are the product of an industry-wide
effort to raise the abstraction level of business systems
development. The Meta-Object Facility (MOF) is a set of standardized
interfaces, including the XML Metadata Interchange format (XMI), that
provide the basis for the specification models required for MDA.  A
Platform Independent Model (PIM) specifies application features
generically that are converted by rule-driven translators into
Platform Specific Models (PSM) that reflect the unique properties of
disparate platforms.  A PSM can be either directly interpreted or
further processed to generate a deployable system.  Web applications
are supported by MDA tools as another implementation platform to
target for code generation.  Large vendors are backing the MDA
standards with compliant toolsets, including Oracle ADF and
IBM/Rational Rapid Developer, and comprehensively support Web
application and Web service development.  While MDA has the potential
to shield developers from the implementation complexities inherent in
Web applications and improving the process, the MDA processes
represent a major paradigm shift for organizations and widespread
diffusion of the technologies, if is occurs, will be incremental.

\subsection{Authoring Tools and Development Environments}
A diverse range of development tools supports the creation of Web
applications.  The simplest Web authoring tools support static page
creation through direct editing, WYSIWIG HTML editing, and the ability
to save documents into other formats to an HTML representation.  The
tools in this category include Adobe PageMill, Amaya, and Microsoft
Office applications such as Word, Excel, and PowerPoint.  The next
level of sophistication is found in site management tools that include
HTML editors.  Microsoft FrontPage, NetObjects Fusion, and initial
versions of MacroMedia Dreamweaver are in this category, featuring
tools for managing a site's link structure, simplifying deployment,
and assisting in the introduction of client-based dynamism into pages.
Recent versions of Dreamweaver feature tighter integration with
server-side dynamic content generation technologies, blurring the
distinction between Web authoring tools and full-fledged Web
application development environments.  This latter category includes
Microsoft VisualStudio.NET, Sun Java Studio Creator, and the
Eclipse-based WebSphere Application Developer.  Full-fledged
development environments have all of the functionality of Web
authoring tools, while also supporting programming-intensive
activities.  As the development processes move towards component-based
methods, the exposed APIs of the components may lead to less
code-intensive development processes for dynamic web appliations and
increased convergence of the development enviroments towards the
simplicity of Web authoring tools.

\subsection{Summary}

\begin{table}
\begin{center}
\begin{tabular}{|p{1.1in}||p{1.4in}|p{2.2in}|} \hline
Technology & Exemplars & Key Properties and Common Usage \\ \hline \hline
{\bf Frameworks} & \multicolumn{2}{|l|}{} \\ \hline      
   Scripting language-based & Twisted (Python), WebWare (Python), Snakelets (Python), Rails (Ruby) 
      & Portable. Small to large-scale applications with complex navigation requirements. \\ \hline
   Application-driven & Struts, Spring MVC, WebWork, Maverick, Barracuda 
      & Portable.  Scalable.  Enterprise web sites with complex navigation requirements. \\ \hline
   Page-driven, component-based, desktop model & Echo, wingS, WebCream, Wi.Ser 
      & Portable.  Complex.  Ease transition to Web programming. \\ \hline
   Page-driven  component-based, Web model & WebObjects, Tapestry, ASP.NET WebForms 
      & Portable.  Scalable.  High usability. Rapid development. Simple to 
        medium-complexity navigation requirements. \\ \hline
   Application/Page Hybrid-driven component-based & JavaServer Faces 
      & Portable.  Scalable.  High usability.  Enterprise web sites with 
        complex navigation requirements.  Rapid development. \\ \hline
   Portal composition & Java Portlet specification, WebParts, Tiles, SiteMesh 
      & Portable. Scalable.  Enterprise portal development. \\ \hline 
{\bf Model Driven Development} & \multicolumn{2}{|l|}{} \\ \hline
   Data-centered & AutoWeb, RMM, OOHDM, Oracle Web Development Suite, WebML, WebRatio
      & Data-intensive applications.  \\ \hline
   GUI-centered & CodeCharge, CodeSmith, DeKlarit, Fabrique
      & Interation-intensive applications. \\ \hline
   MDA-compliant & Oracle ADF, IBM/Rational Rapid Developer
      & OMG standard. \\ \hline
{\bf Development Environments} & \multicolumn{2}{|l|}{} \\ \hline
   Authoring tools & Adobe PageMill, Amaya, Microsoft Office (Word, Excel, PowerPoint)
      & Static content creation and editing.  \\ \hline
   Web site management & FrontPage, NetObjects Fusion, MacroMedia Dreamweaver
      & Mostly static web sites with simple navigation requirements.\\ \hline
   Full-fledged environments & Microsoft VisualStudio.NET, Sun Java Studio Creator, WebSphere Application Developer
      & Programming-intensive. \\ \hline
{\bf Others} & \multicolumn{2}{|l|}{} \\ \hline  
   Programming languages & MAWL, $<$bigwig$>$, JWIG, functional continuations, Cocoon Flow 
      & Research projects. \\ \hline 
\end{tabular}
\caption{Server-side development approaches.}
\label{tab:serverframeworks}
\end{center}
\end{table}

Table \ref{tab:serverframeworks} presents a summary of the current
server-side application development approaches. In spite of the amount
of effort that has been focused on bringing discipline to web
programming, as evidenced by number of approaches that have been
tried, Web application development is still hard. Even when using the
most advanced frameworks or development environments to program simple
applications, programmers must internalize several programming modes
and languages within a single implementation stream. Examples are
abundant, as in the case of the .NET Web programmer who must be
minimally be familiar with HTTP, C$^{\# }$, ASP.NET, JavaScript, HTML,
XML, WebForms, VisualStudio.NET, and various middleware interfaces of
the .NET Framework, including ADO.NET for access to databases, before
even starting to build an application.  The mental models are complex,
not intuitive, and have built-in conceptual barriers that extend
beyond the details of particular technologies.  Newcomers face a long
learning curve before they can become productive.

\section{CONCLUSIONS}
\label{sec:conclusions}

We have presented an extensive survey and several categorizations of
technologies related to web programming. The several tables presented
throughout the paper can serve as a roadmap for choosing technical
artifacts according to a specific application needs. The conclusions
of our study can be summarized as follows.

\noindent
{\bf Infrastructure.}  The technical problems related to the web
infrastructure have largely been solved. Scalability is simply a
matter of cost, as evidenced by the existence of Google which serves
over 1000 users per second with sub-second average response time.
High levels of scalability are achieved through load balancing between
peer systems, and clustering systems are used to balance the
processing load between clusters of low-cost servers. While the Google
solution is beyond the technical capabilities of most organizations,
the load balancing features of J2EE and .NET provide practical
scalability.  These enterprise solutions also have features that can
assure reasonable levels of reliability, extensibility, portability,
and security of Web applications.

\noindent
{\bf Application Development.}  Simple applications that provide only
small amounts of dynamic content are appropriately supported by ad-hoc
programming such as CGI. However, as applications become interaction-
and data-intensive, those ad-hoc programming methods don't scale
well. Industrial-strength application development platforms such as
J2EE and .NET, built on several years of ad-hoc experimentation, aim
at simplifying application development. Unfortunately, the learning
curve for mastering one of these platforms is long and steep, and even
the most proficient software engineer who can master non-web
application development will be overwhelmed by the complexity. Also,
support for the coordinated separation between designers, business
experts and programmers is lacking. The most interesting challenges
ahead lie in effectively simplifying the processes and methods of web
application development.

In the last few years, a significant effort has been devoted to
devising methods that exploit the technology base in a disciplined
way. These methods include frameworks based on proven design patterns,
experimental programming language approaches, model-driven
architectures and development environments. Although promising, they
carry along preconceptions brought from non-Web application
development. It is too early to assess their suitability and
impact. Overall, our study led us to believe that the most critical
element at this point is to formulate a concise and simple model of
what these applications are about, and to build a programming system
around such a model.

\bibliographystyle{acmtrans}
\bibliography{web_programming_survey}

\begin{thebibliography}{}

\bibitem[\protect\citeauthoryear{Abdelnur and Hepper}{Abdelnur and
  Hepper}{2003}]{Abdelnur:web:2003}
{\sc Abdelnur, A.} {\sc and} {\sc Hepper, S.} 2003.
\newblock {JSR} 168: Portlet specification.
\newblock JCP specification. \url{http://www.jcp.org/en/jsr/detail?id=168}.

\bibitem[\protect\citeauthoryear{ActiveState}{ActiveState}{2003}]{ActiveState:%
manual:2003}
ActiveState 2003.
\newblock {\em PerlEx - Online Docs}.
\newblock ActiveState.
\newblock ~\url{http://aspn.activestate.com/ASPN/docs/ASPNTOC-PERLEX____/}.

\bibitem[\protect\citeauthoryear{Altheim and McCarron}{Altheim and
  McCarron}{2001}]{Altheim:web:2001}
{\sc Altheim, M.} {\sc and} {\sc McCarron, S.} 2001.
\newblock {XHTML} 1.1 - module-based {XHTML}: {W3C} recommendation 31 {May}
  2001.
\newblock W3C Recommendation. \url{http://www.w3.org/TR/xhtml11/}.

\bibitem[\protect\citeauthoryear{Apache Software Foundation}{Apache Software
  Foundation}{2003}]{asf:manual:2003}
Apache Software Foundation 1996--2003.
\newblock {\em Apache HTTP Server Version 1.3: Apache API Notes}.
\newblock Apache Software Foundation.
\newblock ~\url{http://httpd.apache.org/docs/misc/API.html}.

\bibitem[\protect\citeauthoryear{Apache Software Foundation}{Apache Software
  Foundation}{2004}]{Apache:web:2004}
Apache Software Foundation 1999-2004.
\newblock {\em Apache {HTTP} Server Project}.
\newblock Apache Software Foundation.
\newblock \url{http://httpd.apache.org/}.

\bibitem[\protect\citeauthoryear{Apte, Hansen, and Reeser}{Apte
  et~al\mbox{.}}{2003}]{Apte:cc:2003}
{\sc Apte, V.}, {\sc Hansen, T.}, {\sc and} {\sc Reeser, P.} 2003.
\newblock Performance comparison of dynamic web platforms.
\newblock {\em Computer Communications\/}~{\em 26,\/}~8, 888--898.

\bibitem[\protect\citeauthoryear{Atzeni, Mecca, and Merialdo}{Atzeni
  et~al\mbox{.}}{1997}]{Atzeni:vldb:1997}
{\sc Atzeni, P.}, {\sc Mecca, G.}, {\sc and} {\sc Merialdo, P.} 1997.
\newblock To weave the web.
\newblock In {\em 23rd International Conference on Very Large Databases
  (VLDB'97)}.

\bibitem[\protect\citeauthoryear{Bell}{Bell}{1997}]{Bell:cacm:1997}
{\sc Bell, G.} 1997.
\newblock The body electric.
\newblock {\em Communications of the ACM\/}~{\em 40,\/}~2, 30--32.

\bibitem[\protect\citeauthoryear{Bellovin, Cohen, Havrilla, Hernan, King,
  Lanza, Pesante, Pethia, McAllister, Henault, Goodden, Peterson, Finnegan,
  Katano, Smith, and Lowenthal}{Bellovin
  et~al\mbox{.}}{2000}]{Bellovin:web:2000}
{\sc Bellovin, S.~M.}, {\sc Cohen, C.}, {\sc Havrilla, J.}, {\sc Hernan, S.},
  {\sc King, B.}, {\sc Lanza, J.}, {\sc Pesante, L.}, {\sc Pethia, R.}, {\sc
  McAllister, S.}, {\sc Henault, G.}, {\sc Goodden, R.~T.}, {\sc Peterson,
  A.~P.}, {\sc Finnegan, S.}, {\sc Katano, K.}, {\sc Smith, R.~M.}, {\sc and}
  {\sc Lowenthal, R.~A.} 2000.
\newblock Results of the security in {ActiveX} workshop, {CERT} {Coordination
  Center}, {Pittsburgh, PA}, {August} 22-23, 2000.
\newblock \url{www.cert.org/reports/activeX_report.pdf}.

\bibitem[\protect\citeauthoryear{Berners-Lee}{Berners-Lee}{1989}]{Berners-Lee:%
web:1989}
{\sc Berners-Lee, T.} 1989.
\newblock Information management: A proposal.
\newblock \url{http://www.w3.org/History/1989/proposal.html}.

\bibitem[\protect\citeauthoryear{Berners-Lee, Cailliau, Luotonen, Nielsen, and
  Secret}{Berners-Lee et~al\mbox{.}}{1994}]{Berners-Lee:cacm:1994}
{\sc Berners-Lee, T.}, {\sc Cailliau, R.}, {\sc Luotonen, A.}, {\sc Nielsen,
  H.~F.}, {\sc and} {\sc Secret, A.} 1994.
\newblock The {W}orld-{W}ide {W}eb.
\newblock {\em Communications of the ACM\/}~{\em 37,\/}~8, 76--82.

\bibitem[\protect\citeauthoryear{Berners-Lee, Fielding, and
  Masinter}{Berners-Lee et~al\mbox{.}}{1998}]{Berners-Lee:web:1998}
{\sc Berners-Lee, T.}, {\sc Fielding, R.}, {\sc and} {\sc Masinter, L.} 1998.
\newblock {RFC} 2396: Uniform resource identifiers ({URI}): Generic syntax.
\newblock \url{http://www.ietf.org/rfc/rfc2396.txt}.

\bibitem[\protect\citeauthoryear{Berners-Lee and Fischetti}{Berners-Lee and
  Fischetti}{2000}]{Berners-Lee:book:2000}
{\sc Berners-Lee, T.} {\sc and} {\sc Fischetti, M.} 2000.
\newblock {\em Weaving the Web}.
\newblock HarperCollins, San Francisco, CA.

\bibitem[\protect\citeauthoryear{Bos, \c{C}elik, Hickson, and Lie}{Bos
  et~al\mbox{.}}{2004}]{Bos:web:2004}
{\sc Bos, B.}, {\sc \c{C}elik, T.}, {\sc Hickson, I.}, {\sc and} {\sc Lie,
  H.~W.} 2004.
\newblock Cascading style sheets, level 2 revision 1, {CSS} 2.1 specification:
  {W3C} candidate recommendation 25 {F}ebruary 2004.
\newblock W3C Recommendation. \url{http://www.w3c.org/TR/CSS21/}.

\bibitem[\protect\citeauthoryear{Brown}{Brown}{1996}]{Brown:web:1996}
{\sc Brown, M.~R.} 1996.
\newblock {FastCGI}: A high-performance gateway interface.
\newblock Position paper for the workshop ``Programming the Web - a search for
  APIs'', Fifth International World Wide Web Conference, 6 May 1996, Paris,
  France.
  \url{http://www.fastcgi.com/devkit/doc/fastcgi-whitepaper/fastcgi.htm}.

\bibitem[\protect\citeauthoryear{Bryant}{Bryant}{2004}]{Bryant:web:2004}
{\sc Bryant, A.} 2004.
\newblock Tutorial: A walk on the {Seaside}.
\newblock \url{http://beta4.com/seaside2/tutorial.html}.

\bibitem[\protect\citeauthoryear{Cardelli and Davies}{Cardelli and
  Davies}{1999}]{Cardelli:ieeetse:1999}
{\sc Cardelli, L.} {\sc and} {\sc Davies, R.} 1999.
\newblock Service combinators for web computing.
\newblock {\em IEEE Transactions on Software Engineering\/}~{\em 25,\/}~3,
  309--316.

\bibitem[\protect\citeauthoryear{Clark}{Clark}{1999}]{Clark:web:1999}
{\sc Clark, J.} 1999.
\newblock {XSL} transformations ({XSLT}) version 1.0: {W3C} recommendation 16
  {N}ovember 1999.
\newblock W3C Recommendation. \url{http://www.w3.org/TR/xslt}.

\bibitem[\protect\citeauthoryear{CreamTec, LLC}{CreamTec,
  LLC}{2005}]{Creamtec:manual:2005}
CreamTec, LLC 2005.
\newblock {\em {WebCream} White Paper}.
\newblock CreamTec, LLC.
\newblock \url{http://www.creamtec.com/webcream/}.

\bibitem[\protect\citeauthoryear{Esposito}{Esposito}{2003}]{Esposito:web:2003}
{\sc Esposito, D.} 2003.
\newblock The {ASP.NET} page object model: One day in the life of an {ASP.NET}
  web page.
\newblock {\em MSDN Library\/}.
\newblock
  \url{http://msdn.microsoft.com/library/default.asp?url=/library/en-us/dnaspp%
/html/aspnet-pageobjectmodel.asp}.

\bibitem[\protect\citeauthoryear{Fayad and Schmidt}{Fayad and
  Schmidt}{1997}]{Fayad:cacm:1997}
{\sc Fayad, M.~E.} {\sc and} {\sc Schmidt, D.~C.} 1997.
\newblock Object-oriented application frameworks: Introduction.
\newblock {\em Communications of the ACM\/}~{\em 40,\/}~10, 32--38.

\bibitem[\protect\citeauthoryear{Felleisen}{Felleisen}{2002}]{conf:afp:Felleis%
en:2002}
{\sc Felleisen, M.} 2002.
\newblock Developing interactive web programs.
\newblock In {\em Advanced Functional Programming}, {J.~Jeuring} {and}
  {S.~L.~P. Jones}, Eds. Lecture Notes in Computer Science, vol. 2638.
  Springer, 100--128.

\bibitem[\protect\citeauthoryear{Fernandez, Suciu, and Tatarinov}{Fernandez
  et~al\mbox{.}}{1999}]{Fernandez:dsl:1999}
{\sc Fernandez, M.~F.}, {\sc Suciu, D.}, {\sc and} {\sc Tatarinov, I.} 1999.
\newblock Declarative specification of data-intensive web sites.
\newblock In {\em Domain-Specific Languages (DSL)}. 135--148.

\bibitem[\protect\citeauthoryear{Fielding, Gettys, Mogul, Frystyk, Masinter,
  Leach, and Berners-Lee}{Fielding et~al\mbox{.}}{1999}]{Fielding:web:1999}
{\sc Fielding, R.}, {\sc Gettys, J.}, {\sc Mogul, J.}, {\sc Frystyk, H.}, {\sc
  Masinter, L.}, {\sc Leach, P.}, {\sc and} {\sc Berners-Lee, T.} 1999.
\newblock {RFC} 2616: Hypertext transfer protocol -- {HTTP}/1.1.
\newblock W3C Recommendation. \url{ftp://ftp.isi.edu/in-notes/rfc2616.txt}.

\bibitem[\protect\citeauthoryear{Fielding and Taylor}{Fielding and
  Taylor}{2002}]{Fielding:acmtoit:2002}
{\sc Fielding, R.~T.} {\sc and} {\sc Taylor, R.~N.} 2002.
\newblock Principled design of the modern web architecture.
\newblock {\em ACM Transactions on Internet Technology\/}~{\em 2,\/}~2,
  115--150.

\bibitem[\protect\citeauthoryear{Fowler}{Fowler}{2002}]{Fowler:book:2002}
{\sc Fowler, M.} 2002.
\newblock {\em Patterns of Enterprise Application Architecture}.
\newblock Addison-Wesley, Boston, MA.

\bibitem[\protect\citeauthoryear{Fraternali}{Fraternali}{1999}]{Fraternali99:a%
cmcsur:1999}
{\sc Fraternali, P.} 1999.
\newblock Tools and approaches for developing data-intensive web applications:
  A survey.
\newblock {\em ACM Computing Surveys\/}.

\bibitem[\protect\citeauthoryear{Garrett}{Garrett}{2005}]{Garrett:web:2005}
{\sc Garrett, J.~J.} 2005.
\newblock Ajax: A new approach to {Web} applications.
\newblock
  \url{http://www.adaptivepath.com/publications/essays/archives/000385print.ph%
p}.

\bibitem[\protect\citeauthoryear{Gellersen and Gaedke}{Gellersen and
  Gaedke}{1999}]{Gellersen:ieeeic:1999}
{\sc Gellersen, H.-W.} {\sc and} {\sc Gaedke, M.} 1999.
\newblock Object-oriented web application development.
\newblock {\em IEEE Internet Computing\/}~{\em 3,\/}~1, 60--68.

\bibitem[\protect\citeauthoryear{Gousios and Spinellis}{Gousios and
  Spinellis}{2002}]{conf:sane:Gousios:2002}
{\sc Gousios, G.} {\sc and} {\sc Spinellis, D.} 2002.
\newblock A comparison of portable dynamic web content technologies for the
  apache web server.
\newblock In {\em Proceedings of the 3rd International System Administration
  and Networking Conference SANE 2002}. 103--119.

\bibitem[\protect\citeauthoryear{Graham}{Graham}{2001}]{Graham:web:2001}
{\sc Graham, P.} 2001.
\newblock Lisp in web-based applications.
\newblock
  \url{http://lib1.store.vip.sc5.yahoo.com/lib/paulgraham/bbnexcerpts.txt}.

\bibitem[\protect\citeauthoryear{Hassan and Holt}{Hassan and
  Holt}{2000}]{conf:wcre:Hassan:2000}
{\sc Hassan, A.~E.} {\sc and} {\sc Holt, R.~C.} 2000.
\newblock A reference architecture for web servers.
\newblock In {\em WCRE}. 150--159.

\bibitem[\protect\citeauthoryear{Hunter}{Hunter}{2000}]{Hunter:web:2000}
{\sc Hunter, J.} 2000.
\newblock The problems with {JSP}.
\newblock \url{http://www.servlets.com/soapbox/problems-jsp.html}.

\bibitem[\protect\citeauthoryear{Husted, Dumoulin, Franciscus, and
  Winterfeldt}{Husted et~al\mbox{.}}{2003}]{Husted:book:2003}
{\sc Husted, T.}, {\sc Dumoulin, C.}, {\sc Franciscus, G.}, {\sc and} {\sc
  Winterfeldt, D.} 2003.
\newblock {\em Struts in Action}.
\newblock Manning Publications, Greenwich, CT.

\bibitem[\protect\citeauthoryear{Johnson and Hoeller}{Johnson and
  Hoeller}{2004}]{Johnson:book:2004}
{\sc Johnson, R.} {\sc and} {\sc Hoeller, J.} 2004.
\newblock {\em Expert One-on-One J2EE Development without EJB}.
\newblock Wiley, Indianapolis, IN.

\bibitem[\protect\citeauthoryear{Krasner and Pope}{Krasner and
  Pope}{1988}]{Krasner:joop:1988}
{\sc Krasner, G.~E.} {\sc and} {\sc Pope, S.~T.} 1988.
\newblock A cookbook for using the model-view controller user interface
  paradigm in smalltalk-80.
\newblock {\em Journal of Object Oriented Program.\/}~{\em 1,\/}~3, 26--49.

\bibitem[\protect\citeauthoryear{Mann}{Mann}{2004}]{Mann:book:2004}
{\sc Mann, K.} 2004.
\newblock {\em JavaServer Faces in Action}.
\newblock Manning Publications, Greenwich, CT.

\bibitem[\protect\citeauthoryear{Manola}{Manola}{1999}]{Manola:ieeeic:1999}
{\sc Manola, F.} 1999.
\newblock Technologies for a web object model.
\newblock {\em IEEE Internet Computing\/}~{\em 3,\/}~1, 38--47.

\bibitem[\protect\citeauthoryear{Meijer}{Meijer}{2000}]{Meijer:jfp:2000}
{\sc Meijer, E.} 2000.
\newblock Server side web scripting in haskell.
\newblock {\em Journal of Functional Programming\/}~{\em 10,\/}~1, 1--18.

\bibitem[\protect\citeauthoryear{Microsoft Corporation}{Microsoft
  Corporation}{2005a}]{Microsoft:web:2005:b}
Microsoft Corporation 2005a.
\newblock {\em IIS Web Development SDK}.
\newblock Microsoft Corporation.
\newblock
  \url{http://msdn.microsoft.com/library/default.asp?url=/library/en-us/iissdk%
/html/e8eff418-8e4f-4db8-ad70-01473bf10741.asp}.

\bibitem[\protect\citeauthoryear{Microsoft Corporation}{Microsoft
  Corporation}{2005b}]{Microsoft:web:2005}
Microsoft Corporation 2005b.
\newblock {\em Internet {I}nformation {S}ervices}.
\newblock Microsoft Corporation.
\newblock \url{http://www.microsoft.com/windowsserver2003/iis/}.

\bibitem[\protect\citeauthoryear{Nash}{Nash}{2003}]{Nash:book:2003}
{\sc Nash, M.} 2003.
\newblock {\em Java Frameworks and Components: Accelerate Your Web Application
  Development}.
\newblock Cambridge University Press, Cambridge, UK.

\bibitem[\protect\citeauthoryear{NCSA}{NCSA}{1993}]{NCSA:web:1993}
NCSA 1993.
\newblock {\em The Common Gateway Interface}.
\newblock NCSA.
\newblock \url{http://hoohoo.ncsa.uiuc.edu/cgi/}.

\bibitem[\protect\citeauthoryear{NCSA}{NCSA}{1995}]{NCSA:web:1995}
NCSA 1995.
\newblock {\em NCSA Mosaic Common Client Interface}.
\newblock NCSA.
\newblock
  \url{http://archive.ncsa.uiuc.edu/SDG/Software/XMosaic/CCI/cci-spec.html}.

\bibitem[\protect\citeauthoryear{NextApp}{NextApp}{2005}]{Nextapp:manual:2005}
NextApp 2005.
\newblock {\em {Echo} White Paper}.
\newblock NextApp.
\newblock \url{http://www.nextapp.com/products/echo/doc/white_paper.html}.

\bibitem[\protect\citeauthoryear{Ousterhout}{Ousterhout}{1998}]{Ousterhout:iee%
ec:1998}
{\sc Ousterhout, J.~K.} 1998.
\newblock Scripting: Higher-level programming for the 21st century.
\newblock {\em IEEE Computer\/}~{\em 31,\/}~3, 23--30.

\bibitem[\protect\citeauthoryear{{PEAK}}{{PEAK}}{2005}]{PEAK:web:2005}
{PEAK} 2005.
\newblock {\em Python Enterprise Application Kit project homepage.}
\newblock {PEAK}.
\newblock \url{http://peak.telecommunity.com/}.

\bibitem[\protect\citeauthoryear{Pressman}{Pressman}{2000}]{Pressman:ieeesw:20%
00}
{\sc Pressman, R.~S.} 2000.
\newblock Manager - what a tangled web we weave.
\newblock {\em IEEE Software\/}~{\em 17,\/}~1.

\bibitem[\protect\citeauthoryear{Raggett}{Raggett}{1999}]{Raggett:web:1999}
{\sc Raggett, D.} 1999.
\newblock {HTML} 4.01 specification: {W3C} recommendation 24 {D}ecember 1999.
\newblock W3C Recommendation. \url{http://www.w3.org/TR/html4/}.

\bibitem[\protect\citeauthoryear{Raggett, Lam, Alexander, Kmiec, and
  Alexander}{Raggett et~al\mbox{.}}{1997}]{Raggett:book:1997}
{\sc Raggett, D.}, {\sc Lam, J.}, {\sc Alexander, I.~F.}, {\sc Kmiec, M.}, {\sc
  and} {\sc Alexander, I.} 1997.
\newblock {\em Raggett on HTML 4\/}, 2nd Edition ed.
\newblock Addison-Wesley, Reading, MA.

\bibitem[\protect\citeauthoryear{Reilly}{Reilly}{2000}]{Reilly:book:2000}
{\sc Reilly, D.~J.} 2000.
\newblock {\em Inside Server-Based Applications}.
\newblock Microsoft Press, Redmond, WA.

\bibitem[\protect\citeauthoryear{Rolsky and Williams}{Rolsky and
  Williams}{2003}]{Rolsky:book:2003}
{\sc Rolsky, D.} {\sc and} {\sc Williams, K.} 2003.
\newblock {\em Embedding Perl in HTML with Mason}.
\newblock O'Reilly, Cambridge, MA.

\bibitem[\protect\citeauthoryear{Roman, Ambler, Jewell, Roman, Jewell, and
  Marinescu}{Roman et~al\mbox{.}}{2001}]{Roman:book:2001}
{\sc Roman, E.}, {\sc Ambler, S.}, {\sc Jewell, T.}, {\sc Roman, E.}, {\sc
  Jewell, T.}, {\sc and} {\sc Marinescu, F.} 2001.
\newblock {\em Mastering Enterprise JavaBeans\/}, 2nd Edition ed.
\newblock Wiley, New York, NY.

\bibitem[\protect\citeauthoryear{Ship}{Ship}{2004}]{Ship:book:2004}
{\sc Ship, H. M.~L.} 2004.
\newblock {\em Tapestry in Action}.
\newblock Manning Publications, Greenwich, CT.

\bibitem[\protect\citeauthoryear{Sinha}{Sinha}{1997}]{Sinha:book:1997}
{\sc Sinha, P.~K.} 1997.
\newblock {\em Distributed Operating Systems: Concepts and Design}.
\newblock IEEE Press, New York, NY.

\bibitem[\protect\citeauthoryear{Smith, Kay, Raab, and Reed}{Smith
  et~al\mbox{.}}{2003}]{conf:c5:Smith:2003}
{\sc Smith, D.~A.}, {\sc Kay, A.~C.}, {\sc Raab, A.}, {\sc and} {\sc Reed,
  D.~P.} 2003.
\newblock Croquet - a collaboration system architecture.
\newblock In {\em C5}. IEEE Computer Society, 2--.

\bibitem[\protect\citeauthoryear{Stein}{Stein}{1998}]{Stein:book:1998}
{\sc Stein, L.}, Ed. 1998.
\newblock {\em Official Guide to Programming with CGI.pm}.
\newblock Wiley, New York, NY.

\bibitem[\protect\citeauthoryear{Sun Microsystems, Inc.}{Sun Microsystems,
  Inc.}{2003}]{Sun:manual:2003}
Sun Microsystems, Inc. 2003.
\newblock {\em Sun ONE Web Server, NSAPI Programmer's Guide, Version 6.1}.
\newblock Sun Microsystems, Inc.
\newblock ~\url{http://docs-pdf.sun.com/817-1835-10/817-1835-10.pdf}.

\bibitem[\protect\citeauthoryear{Thau}{Thau}{1996}]{Thau:cn:1996}
{\sc Thau, R.} 1996.
\newblock Design considerations for the {A}pache {S}erver {API}.
\newblock {\em Computer Networks\/}~{\em 28,\/}~7-11, 1113--1122.

\bibitem[\protect\citeauthoryear{Thompson, Pazandak, Vasudevan, Manola, Palmer,
  Hansen, and Bannon}{Thompson et~al\mbox{.}}{1999}]{Thompson:acmcsur:1999}
{\sc Thompson, C.~W.}, {\sc Pazandak, P.}, {\sc Vasudevan, V.}, {\sc Manola,
  F.}, {\sc Palmer, M.}, {\sc Hansen, G.}, {\sc and} {\sc Bannon, T.~J.} 1999.
\newblock Intermediary architecture: Interposing middleware object services
  between web client and server.
\newblock {\em ACM Computing Surveys\/}~{\em 31,\/}~2es, 14.

\bibitem[\protect\citeauthoryear{Trauring}{Trauring}{2003}]{Trauring:eaij:2003}
{\sc Trauring, A.} 2003.
\newblock Python: Language of choice for {EAI}.
\newblock {\em EAI Journal\/}, 43--45.

\bibitem[\protect\citeauthoryear{Twisted Matrix Laboratories}{Twisted Matrix
  Laboratories}{2005}]{Twisted:web:2005}
Twisted Matrix Laboratories 2005.
\newblock {\em The Twisted Advantage.}
\newblock Twisted Matrix Laboratories.
\newblock \url{http://twistedmatrix.com/services/twisted-advantage}.

\bibitem[\protect\citeauthoryear{Vinoski}{Vinoski}{2004}]{Vinoski:ieeeic:2004}
{\sc Vinoski, S.} 2004.
\newblock Dark matter revisited.
\newblock {\em IEEE Internet Computing\/}~{\em 8,\/}~4, 81--84.

\bibitem[\protect\citeauthoryear{W3C}{W3C}{1997}]{W3C:web:1997}
{\sc W3C}. 1997.
\newblock Integrating object technology and the {Web}.
\newblock W3C. \url{http://www.w3.org/OOP/}.

\bibitem[\protect\citeauthoryear{W3C}{W3C}{2004a}]{W3C:web:2004}
{\sc W3C}. 2004a.
\newblock Document object model ({DOM}) technical reports.
\newblock W3C. \url{http://www.w3c.org/DOM/DOMTR}.

\bibitem[\protect\citeauthoryear{W3C}{W3C}{2004b}]{W3C:web:2004:b}
{\sc W3C}. 2004b.
\newblock {SOAP} specifications: Latest {SOAP} versions.
\newblock W3C. \url{http://www.w3.org/TR/soap/}.

\bibitem[\protect\citeauthoryear{W3C}{W3C}{2005a}]{W3C:web:2005}
{\sc W3C}. 2005a.
\newblock {S}calable {V}ector {G}raphics (svg).
\newblock W3C. \url{http://www.w3.org/Graphics/SVG/}.

\bibitem[\protect\citeauthoryear{W3C}{W3C}{2005b}]{W3C:web:2005:b}
{\sc W3C}. 2005b.
\newblock Synchronized multimedia.
\newblock W3C. \url{http://www.w3.org/AudioVideo/}.

\bibitem[\protect\citeauthoryear{Ward and Hostetter}{Ward and
  Hostetter}{2003}]{Ward:ijwet:2003}
{\sc Ward, S.} {\sc and} {\sc Hostetter, M.} 2003.
\newblock Curl: A language for web content.
\newblock {\em International Journal of Web Engineering and Technology\/}~{\em
  1,\/}~1, 41--62.

\bibitem[\protect\citeauthoryear{Winer}{Winer}{1999}]{Winer:web:1999}
{\sc Winer, D.} 1999.
\newblock {XML-RPC} specification.
\newblock UserLand Software, Inc. \url{http://www.xmlrpc.com/spec}.

\bibitem[\protect\citeauthoryear{wingS Project Team}{wingS Project
  Team}{2005}]{wings:manual:2005}
wingS Project Team 2005.
\newblock {\em Welcome to {wingS}}.
\newblock wingS Project Team.
\newblock \url{http://wings.mercatis.de/tiki-index.php}.

\bibitem[\protect\citeauthoryear{Wu, Wang, and Wilkins}{Wu
  et~al\mbox{.}}{2000}]{conf:scc:Wu:2000}
{\sc Wu, A.~W.}, {\sc Wang, H.}, {\sc and} {\sc Wilkins, D.} 2000.
\newblock Performance comparison of alternative solutions for web-to-database
  applications.
\newblock In {\em Proceedings of Proceedings of the Southern Conference on
  Computing, Hattiesburg, MS, October 26-28, 2000 (SCC 2000)}.
\newblock
  \url{http://rain.vislab.olemiss.edu/~ww1/Slide_Show_Images/SCC_Amanda/SCC_Am%
anda.pdf}.

\bibitem[\protect\citeauthoryear{Yergeau, Bray, Paoli, Sperberg-McQueen, and
  Maler}{Yergeau et~al\mbox{.}}{2004}]{Yergeau:web:2004}
{\sc Yergeau, F.}, {\sc Bray, T.}, {\sc Paoli, J.}, {\sc Sperberg-McQueen,
  C.~M.}, {\sc and} {\sc Maler, E.} 2004.
\newblock {Extensible Markup Language} ({XML}) 1.0 (third edition): {W3C}
  recommendation 04 {F}ebruary 2004.
\newblock W3C Recommendation.
  \url{http://www.w3c.org/TR/2004/REC-xml-20040204/}.

\bibitem[\protect\citeauthoryear{Zeldman}{Zeldman}{2003}]{Zeldman:book:2003}
{\sc Zeldman, J.} 2003.
\newblock {\em Designing with Web Standards}.
\newblock New Riders, Indianapolis, IN.

\end{thebibliography}
\begin{received}
\end{received}

\end{document}